\documentclass[11pt,a4paper]{article} 
\usepackage{epsfig}
\usepackage{psfig}
\usepackage{cite}
\def\3{\ss}

\begin{document}
\def\alphas{$\alpha_s$}
\newcommand{\xbj}        {\mbox{$x$}}
\newcommand{\als}{$\alpha_s$}
\newcommand{\ee}{\mbox{$e^+e^-$}}
\newcommand{\logxp}     {\mbox{$\ln(1/x_{p})$}}
\newcommand{\logxpmax}  {\mbox{$\ln(1/x_{p})_{max}$}}
\newcommand{\la}{\langle}
\newcommand{\ra}{\rangle}
\newcommand{\sleq} {\raisebox{-.6ex}{${\textstyle\stackrel{<}{\sim}}$}}
\newcommand{\sgeq} {\raisebox{-.6ex}{${\textstyle\stackrel{>}{\sim}}$}}
\newcommand{\xp}     {\mbox{${x_p}$}}
\def\E{\mbox{e}^+\mbox{e}^-}
\def\ep{\mbox{e}^+\mbox{p}}
\def\nc{n_{\mathrm{c}}}
\def\nt{n_{\mathrm{t}}}
\def\R{\rho}

                    
\title {\begin{flushright}{\large DESY--99--041}  \end{flushright}
\vspace*{2cm}
\bf\LARGE Measurement of multiplicity and momentum \\
spectra in the current and target regions \\
of the Breit frame in Deep Inelastic Scattering \\
at HERA }
                    
\author{ZEUS Collaboration }
\date{}
\maketitle
\begin{abstract}
\noindent
Charged particle  production
in  neutral current
deep inelastic scattering (DIS) has been studied using the ZEUS detector.
The evolution of the mean multiplicities,
scaled momenta and transverse momenta in $Q^2$ and $x$
for
$ 10 < Q^2 < 5120\ {\rm GeV^2}$  and $x  > 6\times 10^{-4}$ has been
investigated in the current and target fragmentation regions of the
Breit
frame. 
Distributions in the target region, using HERA data for the first time,
are compared to distributions in the current region.
Predictions based on MLLA and LPHD are inconsistent with the data.
\end{abstract}

\pagestyle{plain}
\thispagestyle{empty}
\clearpage
\pagenumbering{Roman}
\begin{center}                                                                                     
{                      \Large  The ZEUS Collaboration              }                               
\end{center}                                                                                       
  J.~Breitweg,                                                                                     
  S.~Chekanov,                                                                                     
  M.~Derrick,                                                                                      
  D.~Krakauer,                                                                                     
  S.~Magill,                                                                                       
  B.~Musgrave,                                                                                     
  J.~Repond,                                                                                       
  R.~Stanek,                                                                                       
  R.~Yoshida\\                                                                                     
 {\it Argonne National Laboratory, Argonne, IL, USA}~$^{p}$                                        
\par \filbreak                                                                                     
  M.C.K.~Mattingly \\                                                                              
 {\it Andrews University, Berrien Springs, MI, USA}
% ...
\newpage                                                                                           
% ...
$^{  37}$ partially supported by the Foundation for German-Russian Collaboration                   
DFG-RFBR \\ \hspace*{3.5mm} (grant no. 436 RUS 113/248/3 and no. 436 RUS 113/248/2)\\              
% ...
                                                           %
                                                           %
\newpage   
                                                           %
                                                           %
\begin{tabular}[h]{rp{14cm}}                                                                       
$^{a}$ &  supported by the Natural Sciences and Engineering Research                               
          Council of Canada (NSERC)  \\                                                            
$^{b}$ &  supported by the FCAR of Qu\'ebec, Canada  \\                                            
$^{c}$ &  supported by the German Federal Ministry for Education and                               
          Science, Research and Technology (BMBF), under contract                                  
          numbers 057BN19P, 057FR19P, 057HH19P, 057HH29P, 057SI75I \\                              
$^{d}$ &  supported by the MINERVA Gesellschaft f\"ur Forschung GmbH, the                          
German Israeli Foundation, and by the Israel Ministry of Science \\                                
$^{e}$ &  supported by the German-Israeli Foundation, the Israel Science                           
          Foundation, the U.S.-Israel Binational Science Foundation, and by                        
          the Israel Ministry of Science \\                                                        
$^{f}$ &  supported by the Italian National Institute for Nuclear Physics                          
          (INFN) \\                                                                                
$^{g}$ &  supported by the Japanese Ministry of Education, Science and                             
          Culture (the Monbusho) and its grants for Scientific Research \\                         
$^{h}$ &  supported by the Korean Ministry of Education and Korea Science                          
          and Engineering Foundation  \\                                                           
$^{i}$ &  supported by the Netherlands Foundation for Research on                                  
          Matter (FOM) \\                                                                          
$^{j}$ &  supported by the Polish State Committee for Scientific Research,                         
          grant No. 115/E-343/SPUB/P03/154/98, 2P03B03216, 2P03B04616,                             
          2P03B10412, 2P03B05315, and by the German Federal Ministry of                            
          Education and Science, Research and Technology (BMBF) \\                                 
$^{k}$ &  supported by the Polish State Committee for Scientific                                   
          Research (grant No. 2P03B08614 and 2P03B06116) \\                                        
$^{l}$ &  partially supported by the German Federal Ministry for                                   
          Education and Science, Research and Technology (BMBF)  \\                                
$^{m}$ &  supported by the Fund for Fundamental Research of Russian Ministry                       
          for Science and Edu\-cation and by the German Federal Ministry for                       
          Education and Science, Research and Technology (BMBF) \\                                 
$^{n}$ &  supported by the Spanish Ministry of Education                                           
          and Science through funds provided by CICYT \\                                           
$^{o}$ &  supported by the Particle Physics and                                                    
          Astronomy Research Council \\                                                            
$^{p}$ &  supported by the US Department of Energy \\                                              
$^{q}$ &  supported by the US National Science Foundation \\                                       
\end{tabular}                                                                                      
                                                           %
\newpage
\setcounter{page}{1}
\pagenumbering{arabic}
\section{Introduction}
\label{sec:intro}
This paper reports the results of a study 
of the properties of the hadronic final
state in neutral current positron-proton 
deep inelastic scattering (DIS). The fragmentation of
the struck quark in DIS is compared to that of the
quarks produced in \ee~annihilation, allowing the concept of
universality of fragmentation in different processes to be tested.
A comparison is also made of the fragmentation 
associated with the 
outgoing struck quark to that in
the target region; the latter is associated with the initial
state radiation from the incoming proton. The data are compared to 
analytical Quantum Chromodynamic (QCD) 
predictions for the momenta 
spectra for both the current and target region. 

The event kinematics of DIS are determined by the negative square of the 
four-momentum of the virtual exchanged boson,
$Q^2\equiv-q^2$, and the Bjorken scaling variable, $x=Q^2/2P\!\cdot\!q$,
where $P$ is the four-momentum of the proton.
In the Quark Parton Model (QPM),
the interacting quark from the proton carries four-momentum $xP.$
The variable $y$, the fractional energy transfer to the proton in its
rest frame, is related to $x$ and $Q^2$ by $y\approx Q^2/xs$, 
where $\sqrt s$ is the positron-proton centre of mass energy.
The invariant mass, $W$, of the hadronic system is related to $x, Q^2$
and the proton mass, $m_p$, by $W^2 = Q^2(1-x)/x+m_p^2.$ At fixed $Q^2$,
the $W^2$ behaviour reflects the $x$ dependence.

\begin{figure}[t]
\begin{center}\mbox{\epsfig{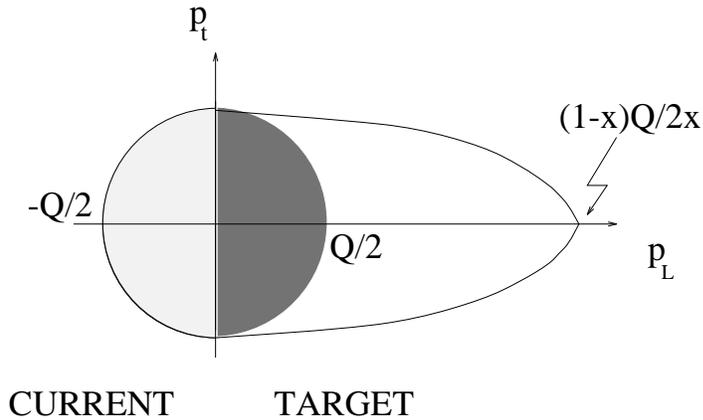}}
\end{center}
\caption{{ The phase space diagram for DIS in the Breit frame,
$p_L$ denotes the longitudinal momentum axis, referred to as the
$z-$direction, and $p_t$ denotes transverse momentum. The dark shaded
region indicates the part of the target region under study in this paper.}}
\label{pic0}
\end{figure}

A natural frame in which to study the dynamics of the hadronic final
state
in DIS is the Breit frame~\cite{feyn}.
In this frame, the exchanged
virtual boson ($\gamma^*$)
is completely space-like and has a four-momentum
$q = (0,0,0,-Q=-2xP^{Breit})\equiv (E,~p_x,~p_y,~p_z)$,
where $P^{Breit}$ is the momentum of the proton in the Breit frame.
The particles produced in the 
interaction can be assigned to one of two regions:
the current region if
their $z$-momentum in the Breit frame is negative, and
the target region if their $z$-momentum is positive (see Fig.~\ref{pic0}).
The main advantage of this
frame is that it gives a
maximal  separation of the incoming and outgoing partons
in the QPM. In this model
the maximum momentum a particle can have in the current region
is $Q/2,$ while in the target region the maximum is $Q(1-x)/2x.$ 
In the Breit frame, unlike the hadronic centre of mass ($\gamma^*p$) frame, 
the two regions
are asymmetric, particularly
at low $x,$ where the target region occupies most of 
the available phase space.

The measurements presented here
extend the previous studies of fragmentation in the
Breit frame performed at HERA~\cite{breit1,breit2,H1breit}.
Increased statistics coming from an integrated luminosity of
$38\ {\rm pb^{-1}}$ 
lead to a significant
improvement in the precision of the scaled momentum distributions
of the charged particles in the current
fragmentation region. 
A subset of data, corresponding to $6.4\ {\rm pb^{-1}},$ has also
been used to measure a small part of the target fragmentation region
(the dark shaded region indicated in Fig.~\ref{pic0}); 
the study of the target region is limited by acceptance and systematic effects 
rather than by statistics. 
The scaled and the transverse momentum
distributions of charged particles in the hadronic final state are
measured in the current and target regions of the Breit frame
as a function of $x$ and $Q^2$ in the ranges
$ 10 < Q^2 < 5120\ {\rm GeV^2}$  and $x  > 6\times 10^{-4}.$ 
Comparisons are made with Monte Carlo
models, QCD analytical calculations and \ee~data.


\section{QCD Models}
The current region in the $ep$ Breit frame
is analogous to a single hemisphere of $e^+e^-$ annihilation.
In $e^+e^- \rightarrow q \bar q$ annihilation the two quarks are produced
with equal and opposite momenta, $\pm \sqrt{s_{ee}}/2.$
The fragmentation of these quarks can be compared to
that of the quark struck from the
proton; this quark has an outgoing momentum $-Q/2$ in the Breit frame.
In the direction of this struck quark 
the scaled momentum spectra of the particles, expressed in terms of
$x_p = 2p^{Breit}/Q,$
are expected~\cite{eedis,anis,char} to have a
dependence on $Q$ similar to that observed in
\ee~annihilation at energy $\sqrt{s_{ee}}=Q,$ with no $x$ dependence.
The effects of higher order processes not present in \ee~annihilation are
discussed in Ref.~\cite{val}.

Results from \ee~annihilation support the need for coherence
effects in perturbative QCD~\cite{bassetto,mueller,webber,MLLA,review}.
The phenomenon of coherence is a natural consequence of the quantum
mechanical
treatment of the parton cascade.
Long wavelength gluons are unable to resolve individual colour
charges
of partons within the parton cascade. This has the effect that
the available phase space for soft
gluon
emissions is reduced to an angular-ordered region, due to destructive
interference. This leads to a number of important differences in the
properties of the partonic final state relative to the incoherent case.
The most notable of these are
a slower rise in the multiplicity of partons with increasing initial
parton energy,
and the modification of the parton logarithmic
momentum spectra to a skewed
Gaussian form, often referred to as the ``hump-backed''
plateau~\cite{MLLA}.
The parton level predictions in practice depend on
 two free parameters, a running 
strong coupling, governed by a QCD scale
$\Lambda,$ and an energy cut-off, $Q_0,$ 
below which the parton
evolution is truncated. 
In this case $\Lambda$ is an effective
scale parameter and is not to be identified with the standard QCD scale,
e.g. $\Lambda_{\rm \overline{MS}}$.
In particular, predictions can be made at
$Q_0 = \Lambda$ yielding the so-called limiting spectrum.

Within the framework of the modified leading log approximation (MLLA)
there are predictions of how the higher order
moments of the parton momentum spectra should evolve 
with the energy scale~\cite{fongweb,dokevol}.
The MLLA calculations are made at the parton level.
The hypothesis of
local parton hadron duality (LPHD)~\cite{LPHD},
which relates the observed hadron
distributions to the calculated parton distributions via a constant of
proportionality, is used in conjunction with the parton predictions
of the MLLA to allow the calculation to be directly compared to data.
There is uncertainty about the energy scale at which the applicability of
LPHD breaks down, in which case the parton predictions cannot be compared to
the hadron distributions.

DIS at low $x$ allows a study of fragmentation in
the target region following the initial scattering off a
sea quark (or antiquark). The
description based on MLLA~\cite{dok} 
is shown schematically
in Fig.~\ref{pic1}, where the quark box at the top of the gluon
ladder represents the scattered sea quark plus its antiquark partner. 
There are various contributions to these calculations.
Contribution C, the top leg of the quark box, corresponds to fragmentation 
in the current region.
Three further contributions (T1, T2 and T3), which are sources of soft gluons,
are considered in these analytical calculations to be associated
with the target region.
It is predicted~\cite{dok} that
the contribution T1
behaves in the same way as the current quark C and so should have
 no $x$ dependence.  
The contribution T2 is due to the colour field
between the remnant and the struck quark,
and the contribution T3 corresponds to
the fragmentation of the rungs in the gluon ladder.
Both T2 and T3 are predicted to have $x$ and
$Q^2$ dependences which differ from T1.
Both the T1 and T2 contributions have been calculated and give
particles of momenta $<Q/2.$ 
The collinear gluons T3, on the other hand,
generally fragment to particles with momentum $\sgeq Q/2.$ 
For values of the scaled momentum $x_p < 1.0,$
the region of phase space 
is analogous to the current region
and has contributions mainly from T1 and T2.
The parton momentum spectra
predicted by MLLA,  over a range of $Q^2$ and $x,$
are shown in more detail in Ref.~\cite{anis}.
In the target region these spectra are approximately 
Gaussian for  $ x_p < 1$;
they peak at a value of $x_p \sim 0.1-0.2$ in the range of $x$ and $Q^2$
measured in this paper, falling to a plateau region 
for  $ 1 < x_p < (1-x)/x$ (the maximum value of $x_p$ in the target region).
The acceptance of the ZEUS detector allows the study of the
contributions from C, from T1 and from T2.

Scaling violations are predicted in the fragmentation functions,
which represent the probability
for a parton to fragment into a particular
hadron carrying a given fraction of the parton's energy.
Fragmentation functions incorporate the long-distance non-perturbative
physics of the hadronisation process in which the observed hadrons are
formed from final state partons of the hard scattering process.
Like parton densities, they
cannot be calculated in perturbative QCD but
can be evolved with the hard-process scale,
using the DGLAP evolution~\cite{DGLAP} equations,
from a starting distribution at a defined energy scale; this starting
distribution can be derived from a fit to data.
If the fragmentation functions are combined with
the cross sections for the inclusive production of
each parton type in the given physical process, predictions can be made
for
scaling violations, expressed as the $Q^2$ evolution 
of the $x_p$ spectra of final state hadrons \cite{webnas}.
These scaling violations
allow a measurement of the strong coupling constant, \alphas,
 and such studies have been
performed at LEP~\cite{ALEPH,DELPHI} by incorporating
lower energy PETRA data.
The NLO calculations (CYCLOPS)~\cite{dirk}
 of the scaled momentum distribution 
exist for DIS but as yet 
the appropriate fragmentation functions at different values of
$\Lambda_{\overline{\rm MS}}$ do not exist to allow the extraction of
$\alpha_s$ from DIS data.

\section{Experimental Setup}
The data presented here were taken at the positron-proton
 collider HERA using the ZEUS detector. The $38\ {\rm pb^{-1}}$
of data 
corresponds to data taken in 1994-1996 and part of the 1997 data sample.
The 1995 data alone, corresponding to $6.4\ {\rm pb^{-1}},$
was used to investigate the target region.
During the period 1994-1997 HERA operated
with positrons of energy $E_e=27.5$~GeV and protons with energy
$820$~GeV.
The ZEUS detector is a multipurpose detector. Of particular importance
in this analysis are the central tracking detector (CTD) and the
uranium-scintillator calorimeter (CAL). 
 A detailed description of the ZEUS detector can be
found in~\cite{b:sigtot_photoprod,b:Detector}.

Throughout
this paper we use the standard ZEUS right-handed coordinate system, in
which
$X = Y = Z = 0$ is the nominal interaction point, the positive
$Z$-axis points in the direction of the proton  beam 
(referred to as the forward direction)
and the $X$-axis is horizontal, pointing towards the centre of HERA.

The tracking system consists of 
a central tracking chamber (CTD)~\cite{b:CTD}
in a 1.43 T solenoidal magnetic field.
The CTD, which surrounds the beampipe,  is a drift chamber consisting of
72~cylindrical layers, arranged in 9 superlayers. Superlayers with
wires parallel to the beam axis alternate with those inclined at a small
angle to give a stereo view.
The single hit efficiency of the CTD is greater than
95$\%.$ 
The resolution of the transverse momentum, $p_t(\rm lab),$ 
in the laboratory frame
for full-length tracks can be parametrised as
$\sigma(p_t({\rm lab}))/p_t({\rm lab})=0.0058p_t({\rm lab}) \oplus 0.0065 
\oplus 0.0014/p_t({\rm lab}),$ with $p_t({\rm lab})$ in GeV.
(This form is a more precise description of the transverse momentum
resolution, particularly at low to medium $p_t({\rm lab})$, 
than that given hitherto).

Surrounding the solenoid is the uranium-scintillator
calorimeter (CAL)~\cite{b:CAL},
which is divided into three parts:
forward, barrel and rear covering the polar regions
$2.6^\circ$ to $36.7^\circ$,
$36.7^\circ$ to $129.1^\circ$ and
$129.1^\circ$ to $176.2^\circ$, respectively.
The CAL covers 99.7$\%$ of the solid angle, with
holes 
in the centres of
the forward and rear calorimeters to
accommodate the HERA beampipe. Each of the calorimeter parts is subdivided
into towers which are segmented longitudinally into electromagnetic (EMC)
and hadronic (HAC) sections. These sections are further subdivided into cells
each of which is read out by two photomultipliers. From test beam data,
energy resolutions
of ${\sigma}_E/E = 0.18/\sqrt{E}$ for electrons and
${\sigma}_E/E = 0.35/\sqrt{E}$ for hadrons ($E$ in~GeV) have been obtained.

The ZEUS detector is almost hermetic, allowing the kinematic
variables $x$ and $Q^2$ to be reconstructed in a variety of
ways using combinations of energies and angles
of the positron and hadronic system.
Variables calculated only
from the measurements of the energy, $E^{\prime}_e,$
and angle, $\theta_e,$
of the scattered positron are denoted with the subscript $e$, whilst
those
calculated from the hadronic system measurements, with the Jacquet
Blondel method~\cite{jb}, are denoted by the subscript $JB.$
Variables calculated by these approaches
are used only in the event selection.
In the double angle method~\cite{DA}, denoted by $DA,$  the
kinematic variables are determined using $\theta_e$
and the angle $ \gamma_H $ (which is the direction of the struck
quark in the QPM), defined from the hadronic
final state.

The $DA$ method was used throughout this analysis
for the calculation of the boosts and the kinematic variables
because it is less
sensitive to systematic
uncertainties in the energy measurement than other methods.

The triggering and online
event selections were identical to those used for the
measurement of the structure function $F_2$~\cite{z_shift}.
The reconstructed tracks used in the charged particle analyses
are associated with the primary event vertex
and have $p_t(\rm lab)>150$~MeV and $|\eta(\rm lab)|<1.75,$
where $\eta(\rm lab)$ is the pseudorapidity 
given by $-\ln(\tan(\theta/2))$ with
$\theta$  being the polar angle of the measured track with respect to
the proton direction in the lab.~frame.
This is the region of CTD acceptance
where the detector response and systematics are best understood.

Further selection
criteria were applied both
to ensure accurate reconstruction of the kinematic
variables and to increase the purity of the sample by
eliminating background from photoproduction processes:

\begin{itemize}
\item
$E^\prime_e \ge 10~{\rm GeV}$,
to achieve a high purity sample of DIS events;
\item
$Q^2_{DA}\geq 10$ GeV$^2,$ to further enhance the purity of the DIS sample; 
\item
$y_e\leq 0.95$,
to reduce the photoproduction background;
\item
$y_{JB}\geq 0.04$, to give sufficient accuracy for $DA$ reconstruction 
of $Q^2$ and $x$;
\item $35 \le \delta = \sum\left( E-p_Z\right)\le 60$~GeV where the
summation is over all calorimeter cells, 
to remove photoproduction events and events with large radiative corrections;
\item $ |X|  >  16\ {\rm cm}\ {\rm or}\ |Y|  >  16\ {\rm cm} ,$ where
  $X$ and $Y$ are the impact position of the
  positron on the CAL,
  to avoid
  the region directly adjacent to the rear beampipe;
\item $ -40  <  Z_{\rm vertex}  <  50\ {\rm cm},$ to reduce background
events from non-$ep$ collisions.
\end{itemize}

The $(x,Q^2)$ bins are listed in Table~\ref{tab:bins}.
 The sizes of the bins were chosen to give good statistics in each
 bin and to limit the migrations between bins~\cite{breit1}.
There is negligible background from non-DIS events.

\section{Event Simulation}
\label{s:model}
Monte Carlo event simulation is used to correct for acceptance and
resolution effects.  The detector simulation is based on the
GEANT~3.13~\cite{GEANT} program and incorporates our best knowledge of
the
apparatus.

To calculate the correction factors, neutral current DIS events
were generated, via the DJANGO~6.24 \mbox{program}~\cite{DJANGO},
using HERACLES~4.5.2 \cite{HERACLES} which
incorporates first order electroweak corrections.
The QCD cascade was modelled
with the colour dipole model, including the boson-gluon fusion process,
using the ARIADNE~4.08 \cite{ariadne} program. In this model coherence
effects are implicitly included in the formalism of the parton cascade.
The colour dipole model treats gluons emitted from
quark--antiquark (diquark) pairs
as radiation from a colour dipole between two partons.
This results in partons that are not ordered in their
transverse momenta.  
The program uses the Lund string fragmentation model \cite{string}
for the hadronisation phase, as implemented
in JETSET~7.4 \cite{JETSET}.
For the analysis of the 1995 data,
two Monte Carlo samples were generated, $4.2~{\rm pb}^{-1}$
with $Q^2>6$ GeV$^2$ and
$15.8~{\rm pb}^{-1}$
with $Q^2>40$ GeV$^2,$ using the
$\rm{GRV94}$ \cite{grv94} parameterisation of the parton distribution
functions.
For the 1996 and 1997 data, a sample with
$Q^2>70$ GeV$^2,$ (with $\rm{MRSA}$ parton densities~\cite{mrsa}), 
was generated, corresponding to $17.1~{\rm pb}^{-1},$
and a sample with 
$Q^2>800$ GeV$^2$ (with $\rm{GRV94}$ parton densities), corresponding to
$53~{\rm pb}^{-1}.$

For the studies of the systematics for the 1995 data,
two additional samples of events were
generated 
($2.1~{\rm pb}^{-1}$ with $Q^2>6$
GeV$^2$ and $9.1~{\rm pb}^{-1}$ with  $Q^2>70$ GeV$^2$)
using the HERWIG~5.8d Monte Carlo program
\cite{herwig},
where no electroweak radiative
corrections were applied.
In HERWIG, coherence effects in the QCD cascades are included by
angular
ordering of successive parton emissions
and a clustering model is used for the hadronisation
\cite{webber,cluster}.
For the 1996 and 1997 data, HERWIG samples with $Q^2>70$ GeV$^2$ 
were generated, corresponding to $9.0~{\rm pb}^{-1},$
and  $Q^2>800$ GeV$^2$ (both with $\rm{MRSA}$ parton densities), 
corresponding to
$60~{\rm pb}^{-1}.$
Both the $\rm{GRV94}$ and $\rm{MRSA}$ parametrisations
agree well with the HERA measurements
of the proton structure function $F_2$ in the $(x,Q^2)$
range of this analysis~\cite{f2,h1f2}.

Another approach to modelling the parton cascade is included in
the LEPTO 6.5.1~\cite{LEPTO}
program, which incorporates the LO $\alpha_s$ matrix element
matched to DGLAP parton showers
(MEPS). This recent version of
LEPTO incorporates the soft colour interaction (SCI) model~\cite{SCI}
to describe HERA rapidity gap events. 
SCI produces changes to the usual
string topologies in non-gap events causing
the string to overlap itself and this results in an increase both of
particle number and energy per unit of rapidity. 

The Linked Dipole
Chain model, LDC 1.0, \cite{LDC}  has also been investigated.
In this model the parton shower evolution is based on
a reformulation \cite{CCFM_LDC} of the CCFM approach \cite{CCFM}
which approximates the BFKL~\cite{BFKL}
prediction at low $x$ and the DGLAP
prediction in the high $x$ limit.
The parton density
parametrisation of ``set A''~\cite{LDC}
was used, which fits data from H1 and ZEUS.
The DGLAP equation predicts strong
ordering of the parton transverse momenta while the BFKL
equation relaxes this ordering but imposes strong ordering of the
longitudinal momenta.
Both the LEPTO and LDC programs use the Lund string fragmentation model.
They were used to compare generator level calculations 
with our data.

\section{Correction Procedure}

The Monte Carlo event samples were used to determine
the mean charged particle acceptance
in the current region as a function of $(x,Q^2)$.
The chosen analysis intervals in $(x,Q^2)$ correspond to regions of
high acceptance (between $74{\ \rm and\ }96\%)$
in the current region of the Breit frame.
The acceptance for the limited area of the target region under
study ($x_p < 1.0$) is lower;
the $\ln(1/x_p)$ distributions have a good acceptance around their
peak positions (70-90\%)  but it falls
to about 50\% at lower values of $\ln(1/x_p)$
for the $(x,Q^2)$ bins 1-4 defined in Table~\ref{tab:bins}. Due to the 
low acceptance and large systematic uncertainties
for $\ln(1/x_p) < 1.0 $ this region
is not studied in  $(x,Q^2)$
bins 5-8 and no studies in the target region are made 
beyond bin 8. There is good acceptance ($>$90\%) for
both regions under study for $p_t > 1.0 {\rm\ GeV,}$ where $p_t$ is
the transverse momentum with respect to the virtual photon direction
in the Breit frame. However the acceptance falls
below 50\% for $p_t < 0.5  {\rm\ GeV.}$
These values are well understood in terms of geometrical acceptances.

\begin{table}[hbt]
\begin{center}
\begin{tabular}{|c|c|c|}
\hline
Bin no. &  $x$ range &  $Q^2$ (GeV$^2$) range  \\
\hline\hline
1 & $(6.0 - 12.0)  10^{-4} $ & $ 10 - 20 $  \\
\hline
2 & $(1.2 - 2.4)  10^{-3} $ & $ 10 - 20 $   \\
3 &    & $ 20 - 40 $    \\
4 &    & $ 40 - 80 $    \\
\hline
5 & $(2.4 - 10.0)  10^{-3} $ & $ 20 - 40 $  \\
6 &    & $ 40 - 80 $   \\
7 &    & $ 80 - 160 $   \\
8 &    & $ 160 - 320 $  \\
\hline
9 & $(1.0 - 5.0)  10^{-2} $ & $ 160 - 320 $  \\
10 &     & $ 320 - 640 $   \\
11 &    & $ 640 - 1280 $   \\
\hline
12 & $0.025 - 0.15  $ & $ 1280 - 2560 $  \\
\hline
13 & $0.05 - 0.25  $ & $ 2560 - 5120 $  \\
\hline\hline
\end{tabular}
\caption{{\small The ($x$,$Q^2$) analysis bins.}}
\label{tab:bins}
\end{center}
\end{table}

About $7\%$ of the tracks generated in the current region
migrate to the target region.
Migrations into the current region from
the target fragmentation region are typically less than 5\% of the
tracks assigned
to the current region for $ Q^2 > 320 {\rm\ GeV^2}.$
For $10 < Q^2 < 320 {\rm\ GeV^2}$
these migrations are on average 12\%,
reaching 25\% for
$Q^2 < 40  {\rm\ GeV^2}$ and
low values of $y$
where the hadronic activity is low
and the measurement of $\gamma_H$ is subject to systematic problems
leading to a worse $x$ resolution
and hence an uncertainty in the boost vector to the Breit frame.

The correction procedure is based on the detailed Monte Carlo
simulation of the ZEUS detector with the event generators
described in the previous section.
Since the ARIADNE model gives the best overall description of our
observed
energy flow~\cite{zeus:efl} it is used for the standard corrections
to the distributions.

The data are corrected for trigger and event selection cuts;
event migration between ($x,Q^2$) intervals;
QED radiative effects;
track reconstruction efficiency;
track selection cuts in $p_t({\rm lab})$ and $\eta({\rm lab})$;
track migration between the current and target regions;
and for the
products of $\Lambda$ and $K^{0}_S$ decays which are assigned to the
primary vertex.

Correction factors were obtained from the Monte Carlo simulation by
comparing the generated distributions, excluding decay
products of $\Lambda$ and $K^0_S,$ with the reconstructed
distributions after the detector and trigger
simulations.
The same reconstruction, selection and analysis were used
for the Monte Carlo simulated events
as for the data. The
correction factors, $F(\xp\!)$, were calculated for each $x_p$~bin 
using a bin-by-bin correction:
$$
F(\xp\!) = \frac{1}{N_{\rm gen}} \; \left( \frac{dn}{d\xp} \right) _{\rm
gen}
 \left/ \frac{1}{N_{\rm obs}} \; \left( \frac{dn}{d\xp} \right) _{\rm
obs}  \right.  $$
where $N_{\rm gen}$ ($N_{\rm obs}$) is the number of generated
(observed)
Monte Carlo events in each $(x,Q^2)$ interval and $n$ is the number of
charged particles (tracks) in the current or target
region in the corresponding $x_p$ and
$(x,Q^2)$ interval.
A similar correction procedure was applied for the other variables.
The bin sizes of the distributions were chosen to be commensurate with
the measurement resolution.
In the current region,
the overall correction factors are greater than unity and
typically $<1.3.$
In the target region these correction factors are larger but, in the region
that we measure, they are
typically $<1.5$ for bins 1-4 and  $<2.0$ for bins 5-8.

\section{Systematic Checks}

The systematic uncertainties
 in the measurement can be divided into three types: 
uncertainties
due to event reconstruction and selection, to track selection, and to 
the Monte Carlo model used.
The systematic checks were as follows:
\begin{itemize}

\item The cut on $y_e<0.95$ was changed to $y_e < 0.8$.   

\item The cut on $ 35 \le \Sigma(\mbox{E - p}_z) \le 60 {\rm\ GeV}$ was changed 
to $40 \le \Sigma(\mbox{E - p}_z) \le  60 {\rm\ GeV}$.

\item The tracking cuts on $|\eta({\rm lab})|<1.75$ and 
$p_t({\rm lab}) > 150 $ MeV were tightened 
to $|\eta({\rm lab})|<1.5$ and 
$p_t({\rm lab}) > 200$ MeV; the cuts were also removed.

\item  Instead of requiring vertex fitted tracks, all reconstructed
tracks that passed through superlayers one and three of the CTD were
accepted.

\item The data were corrected using a different hadronisation model,
namely HERWIG, in place of ARIADNE.

\end{itemize}

With the exception of the change in the model from ARIADNE to HERWIG, 
and the use of non-vertex tracks,
all the systematic effects were small, i.e. within 
two standard deviations of the statistical errors.  Of the two major
systematic uncertainties, the hadronisation model change was dominant.

\newpage
\leftline{\bf Current fragmentation region}

The use of non-vertex tracks resulted in an overall increase in 
the single particle densities of 5\% to
15\% and was fairly flat across the $x_p$ range.
The use of HERWIG to unfold the data gave rise to systematic 
shifts as large
as 15\%. For $Q^2 < 80 \ {\rm GeV^2}$ the tendency of the correction
was to lower the single particle density values 
at low $x_p$ and to increase the values at higher $x_p.$
For $p_t$  in the range $ 0 < p_t < 0.5 {\rm\ GeV,}$ the systematic 
uncertainties 
were about 10\%.
They reduced with increasing $p_t$ to about 5\%.

\vspace*{24.0pt}
\leftline{\bf Target fragmentation region}

Systematic effects due to the different hadronisation models
were largest (as high as 50\%) in the $\ln(1/x_p)$ distributions
at high $x_p.$  
For $x_p < 0.3$ they were typically 10\%, increasing
to 30\% at larger $x_p.$
The tracking systematic from non-vertex tracks was largest at low $x_p$ 
in the lowest $Q^2$ bin where it is 7\%, but otherwise 
was of the order of 2\%.
The $p_t$ distribution was little influenced by the model
used, as would be expected from the good acceptance,
at all but the very lowest transverse momenta. 
The mean $ p_t$ showed a model dependence in the
target region of at most 20\% which was 
due to the reduced acceptance for low $p_t$ tracks
at high $x_p$.
In general, unfolding with HERWIG resulted in a higher value
of the normalised single particle densities.

\section{Results}

\subsection{Current fragmentation region}

\begin{figure}[t]
\centerline{\psfig{figure=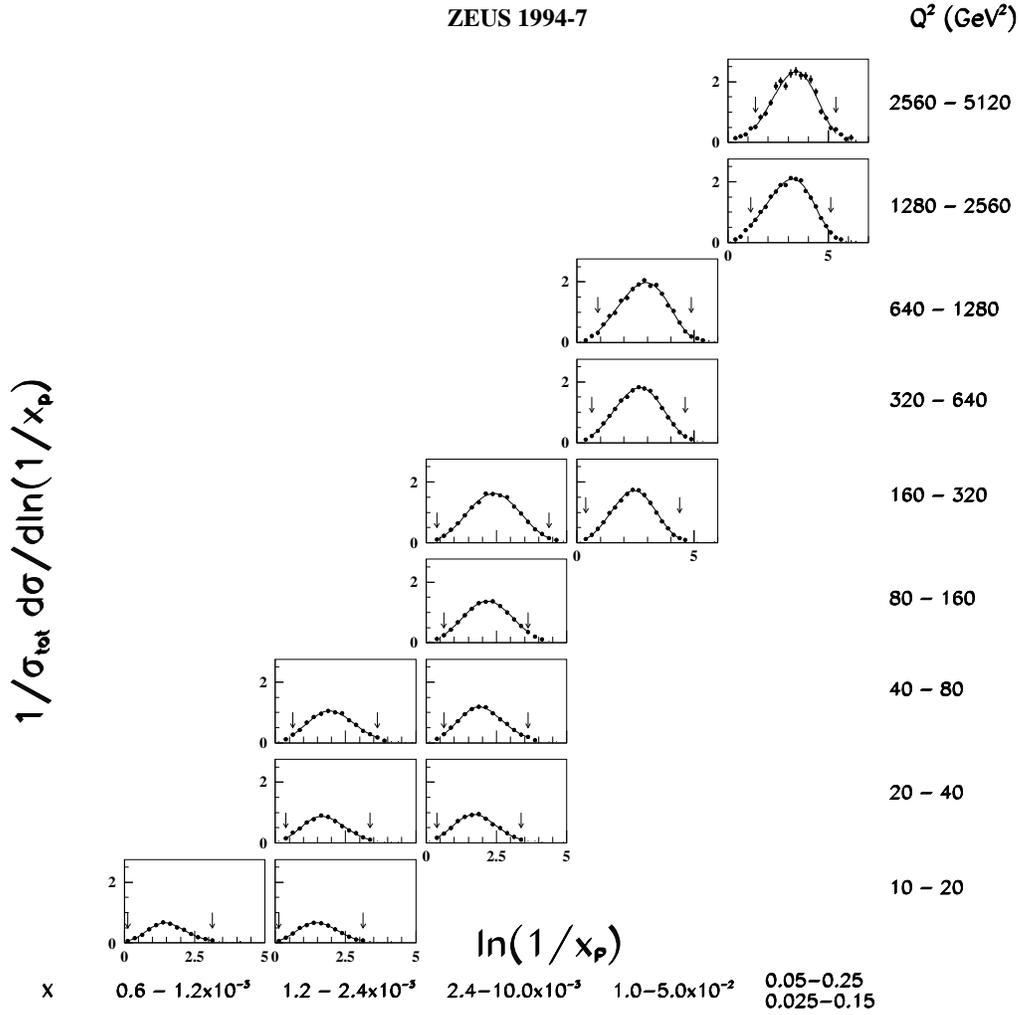,width=0.95\textwidth,
bbllx=15pt,bblly=140pt,bburx=580pt,bbury=705pt}}
\caption{The charged particle distributions of
$1/\sigma_{\rm tot} d\sigma /d\ln(1/x_p)$ in the current fragmentation
region as a function of $\ln(1/x_p)$
for different $(x,Q^2)$ bins.
Only statistical errors are shown. The full line is the skew
Gaussian fit; the arrows indicate the fit range.}
\label{fig:logxp}
\end{figure}

Figure~\ref{fig:logxp} shows the
\logxp~distributions for charged particles in the current fragmentation
region of the Breit frame for different 
bins of $(x,Q^2).$
These distributions are approximately Gaussian in shape with
the mean charged multiplicity given by the integral of the distributions.
As $Q^2$ increases, the multiplicity increases and, in addition,
the peak of the
distribution moves to larger values of \logxp.
The moments of the \logxp\ distributions have been investigated
up to the 4th order;
the mean $(l),$ width $(w),$ skewness $(s)$ and kurtosis $(k)$
 were extracted from each distribution
by fitting a distorted Gaussian of the following form:
\begin{equation}
\frac{1}{\sigma_{tot}} \frac{d\sigma}{d\ln(1/x_p)} \propto
 \exp\left(\frac{1}{8}k-\frac{1}{2}s\delta -\frac{1}{4}(2+k)\delta^2
+\frac{1}{6}s\delta^3 + \frac{1}{24}k\delta^4\right), \label{eq:dg}
\end{equation}
where $\delta = (\ln(1/x_p) - l)/w,$
over a range of $\pm1.5$ units (for $Q^2 < 160 {\rm\ GeV^2}$) or 
$\pm2$ units (for $Q^2 \ge 160 {\rm\ GeV^2}$) in $\ln(1/x_p)$ around the mean.
The equation
is motivated by the expression used for the MLLA predictions of the 
spectra~\cite{fongweb}.
The smooth curves in Fig.~\ref{fig:logxp} result from the fit
of  equation~(\ref{eq:dg}) to the data;
they represent the data well.

\begin{figure}[b]
\centerline{\psfig{figure=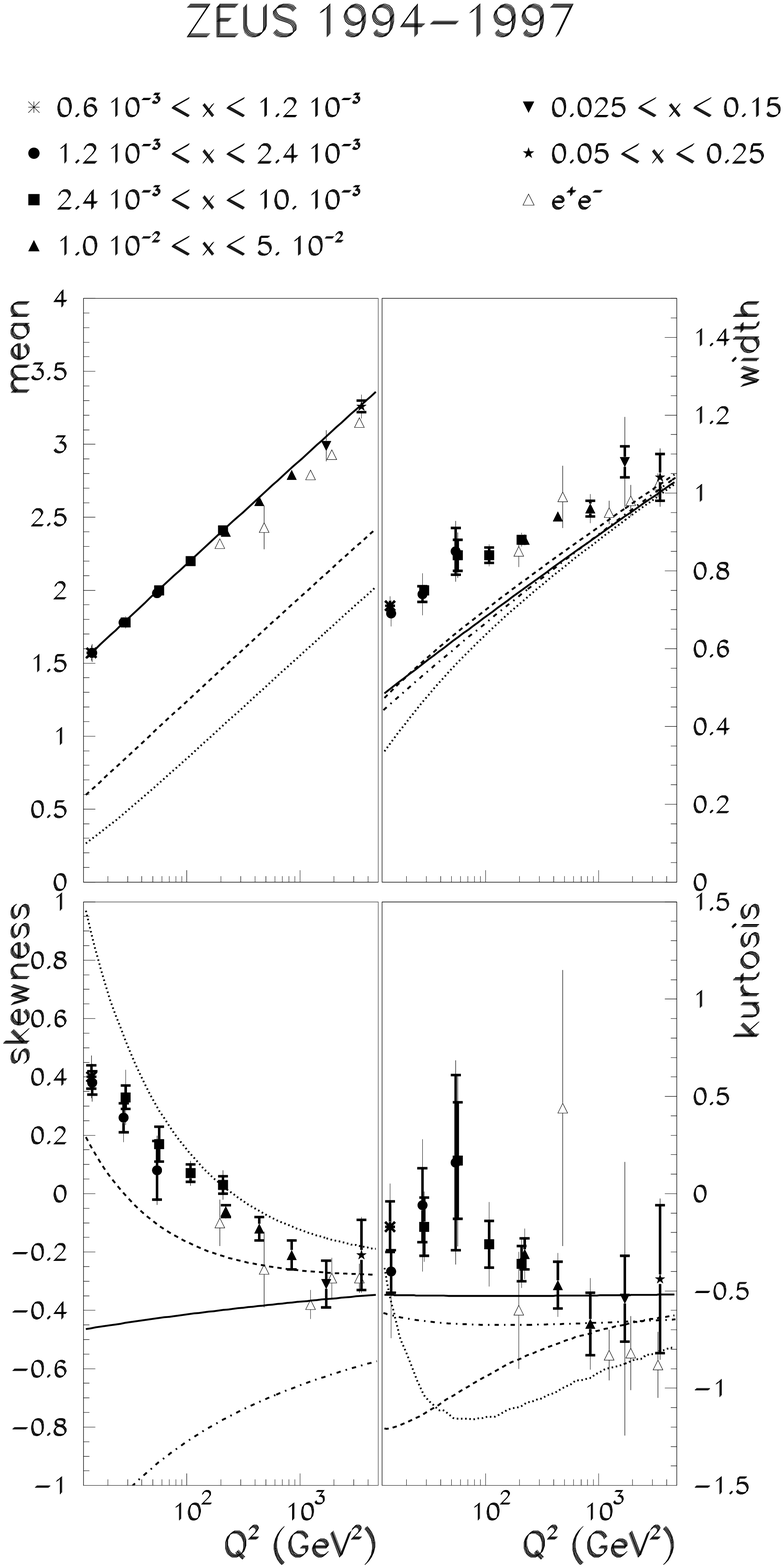,width=\textwidth,height=19.cm}}
\caption{ Evolution of the mean, width, skewness and kurtosis of the
$\ln(1/x_p)$ distribution in the current fragmentation region with $Q^2.$ 
Data from $e^+e^-$ and $ep$ are shown together with 
the MLLA predictions of
Dokshitzer {\it et al}~\protect\cite{dokevol}
(the full line is $Q_0=\Lambda,$ the dashed
$ Q_0= 2\Lambda,$ and the dotted $Q_0 = 3\Lambda$) and
the  limiting spectrum predictions of
Fong and Webber~\protect\cite{fongweb} (dash-dotted line where
available.)
The overlapping points are different $x$ ranges in the same $Q^2$ range.
The
inner error bars are the statistical error and the outer error bars are the
systematic and statistical errors added in quadrature.}
\label{fig:qevol}
\end{figure}

Figure~\ref{fig:qevol} shows the moments of the  $\ln(1/x_p)$
spectra as a function of $Q^2.$
It is evident that the mean and width increase with
increasing $Q^2,$ whereas the skewness and kurtosis decrease. 
Similar fits performed on \ee\ data~\cite{logxpee} 
show a reasonable agreement with our results,
consistent with the universality of fragmentation for this distribution. 

The data are compared to the MLLA predictions of Ref.~\cite{dokevol}, 
using a value of $\Lambda=175{\rm \ MeV},$ 
for different values of $Q_0.$ A comparison is also made with
the predictions of Ref.~\cite{fongweb} for the
limiting spectrum  ($Q_0 = \Lambda$).
The MLLA predictions of the 
limiting spectrum in Ref.~\cite{dokevol}
describe the mean well. However both of the MLLA calculations predict a 
negative skewness which tends towards zero
with increasing $Q^2$ in the case of the limiting spectra.
This is contrary to the measurements. The qualitative description of the
behaviour of the skewness with $Q^2$ can be achieved for a 
truncated cascade ($Q_0 > \Lambda$), but a consistent description of 
the mean, width, skewness and kurtosis
cannot be achieved. 
A range of $\Lambda$ values was investigated and no single value of $\Lambda$
gave a consistent description of all the moments.

We conclude that the
MLLA predictions, assuming LPHD, do not describe the data.
We note however that  a moments analysis has been 
performed~\cite{seroch}, taking into
account the limitations of the  massless assumptions of the MLLA predictions,
and yields good agreement between the
limiting case of the MLLA~\cite{dokevol}
and ${\rm e^+e^-}$ data over a large energy range,
$ 3.0 < \sqrt{s_{ee}} < 133.0\ {\rm GeV}.$
A discussion of phase space effects on the $\ln(1/x_p)$ distributions
is given in ref.~\cite{chlia}. These phase space effects can resemble MLLA.

\begin{figure}[b]
\centerline{\psfig{figure=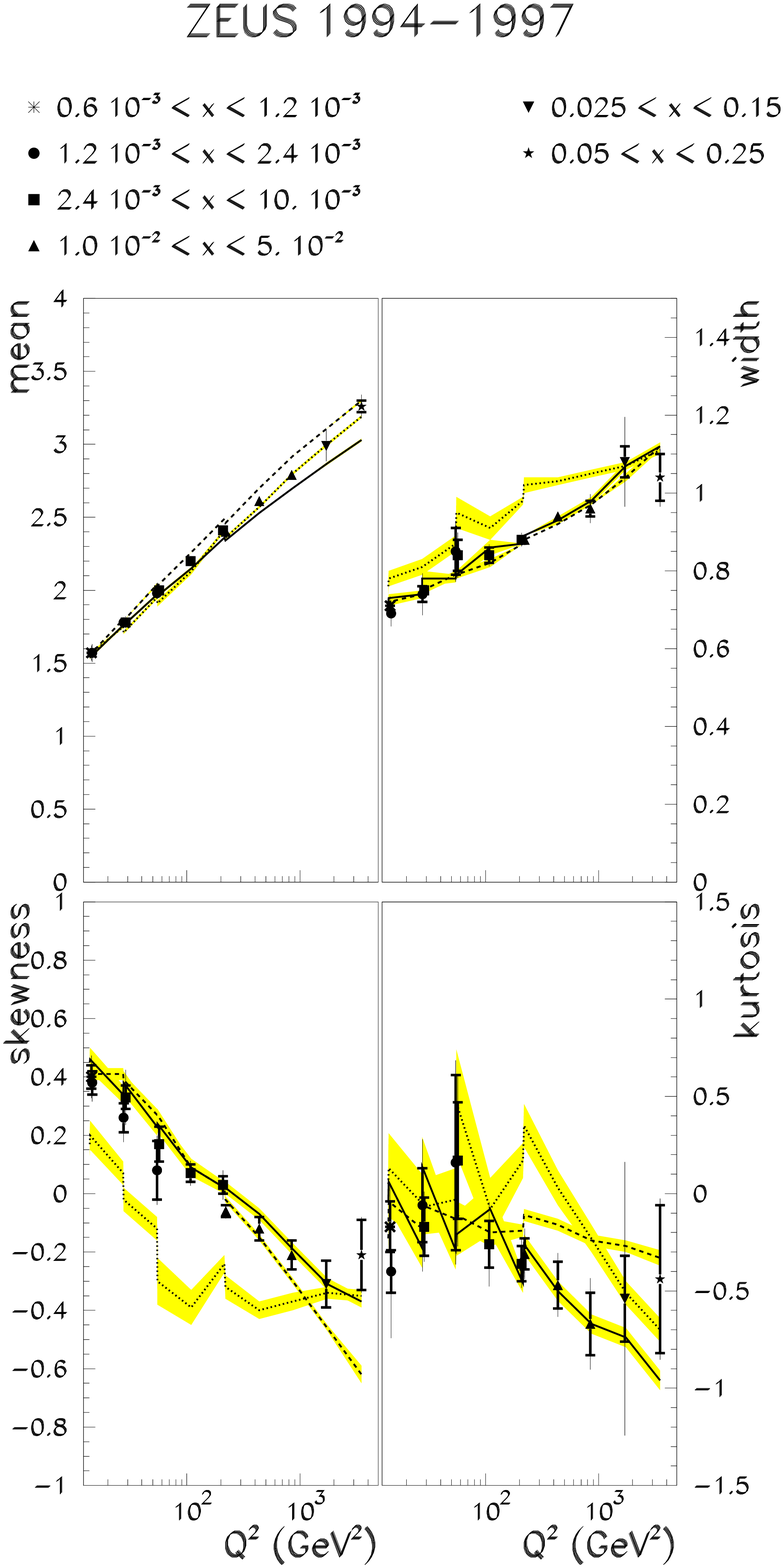,width=\textwidth,height=19.cm}}
\caption{ Evolution of the mean, width, skewness and kurtosis of the
$\ln(1/x_p)$ distribution in the current fragmentation region with $Q^2.$
The overlapping points are for different $x$ ranges in the same $Q^2$ range.
The
inner error bars are the statistical error and the outer error bars are the
systematic and statistical errors added in quadrature. The lines
 are the predictions from 
the Monte Carlo generators ARIADNE (full), LEPTO with SCI (dashed) and HERWIG
(dotted). The shaded region represents the statistical error from the fits
to the Monte Carlo simulations. 
The LEPTO model without SCI resembles the predictions
of ARIADNE.}
\label{fig:qevolmc}
\end{figure}

In Fig.~\ref{fig:qevolmc} the evolution of the moments with $Q^2$
(same DIS data as Fig.~\ref{fig:qevol}) 
are compared with the predictions of various 
Monte Carlo models. Both ARIADNE and LEPTO (with SCI) give a reasonable
description of the data, while HERWIG fails to predict the observed $Q^2$
variation.
This is particularly noticeable for skewness and kurtosis.  
The discontinuities in the HERWIG prediction arise from
a  strong $x$ dependence
in bins of overlapping $Q^2.$ Such an $x$ dependence is
not observed in the data. 
It may be noted that, due to the choice of the
maximum scale of the parton shower evolution, 
there are fewer gluons radiated in HERWIG than in the other generators;
this could possibly
account for the poor agreement of HERWIG with our measurement.
All Monte Carlo programs have been compared using the default values of their 
parameters. The LEPTO model without SCI (not shown) describes the data
better than does the default version.

\begin{figure}[t]
\centerline{\psfig{figure=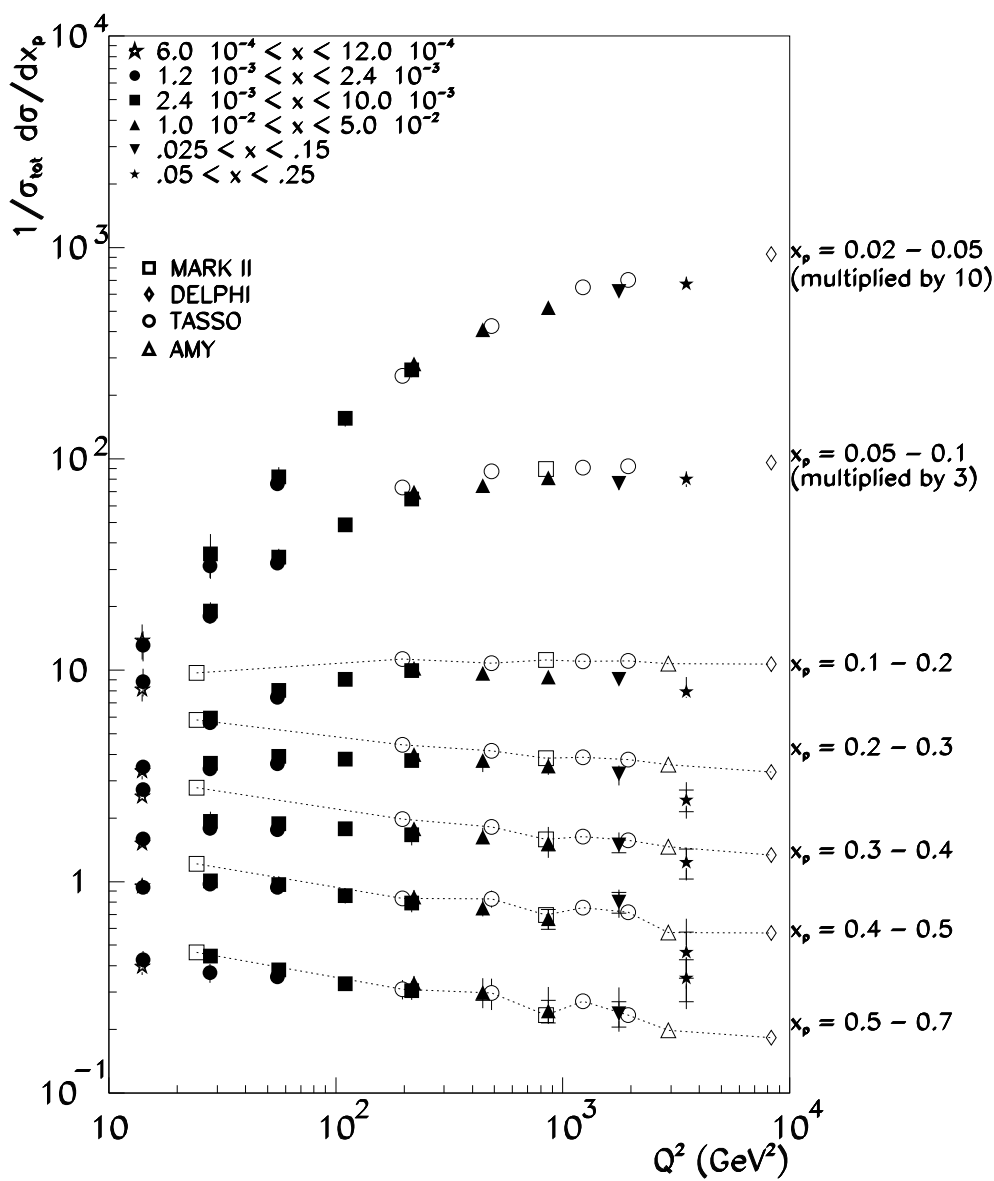,width=0.95\textwidth}}
\caption{The inclusive charged particle distribution,
$ 1/\sigma_{tot}~ d\sigma/dx_p$,
in the current fragmentation region of the Breit frame.
The
inner error bar is the statistical and the outer error bar shows the
systematic and statistical errors added in quadrature.
The open points represent data from $e^+e^-$ experiments divided by two
to
account for $q$ and $\bar q$ production (also corrected
for contributions to the charged multiplicity from $K^0_S$ and $\Lambda$
decays). The low energy MARK II data has been offset slightly to the
left for the sake of clarity.}
\label{fig:largexp}
\end{figure}

The inclusive charged particle distribution,
$ 1/\sigma_{tot}~ d\sigma/dx_p$,
in the current fragmentation region of the Breit frame is shown in bins
of $x_p$ and $Q^2$ in Fig.~\ref{fig:largexp}.
The fall-off as $Q^2$ increases for $x_p > 0.3$ 
(corresponding to  the production of more particles with
a smaller fractional momentum) is indicative of scaling
violations in the fragmentation function.
The distributions rise
with $Q^2$ for $x_p <0.1$ and are discussed in more detail below.
The data are compared
to $e^+e^-$ data~\cite{eedata} 
(divided by two to account for the production of a $q\bar q$ pair) 
at  $Q^2=s_{ee}.$
For the higher $Q^2$ values shown there is a good agreement between the
measurements in the
current region of the Breit frame in DIS and the $e^+e^-$
results; this again supports
the universality of fragmentation. 
The fall-off observed in the ZEUS data at low $x_p$ and low $Q^2$ 
is greater than that observed in $e^+e^-$ data at SPEAR~\cite{spear};
this can be attributed to
processes not present in $e^+e^-$ (e.g. scattering off a sea quark and/or
boson gluon fusion (BGF)) which depopulate the current 
region~\cite{val,diffeflow,eden}.

\begin{figure}[b]
\centerline{\psfig{figure=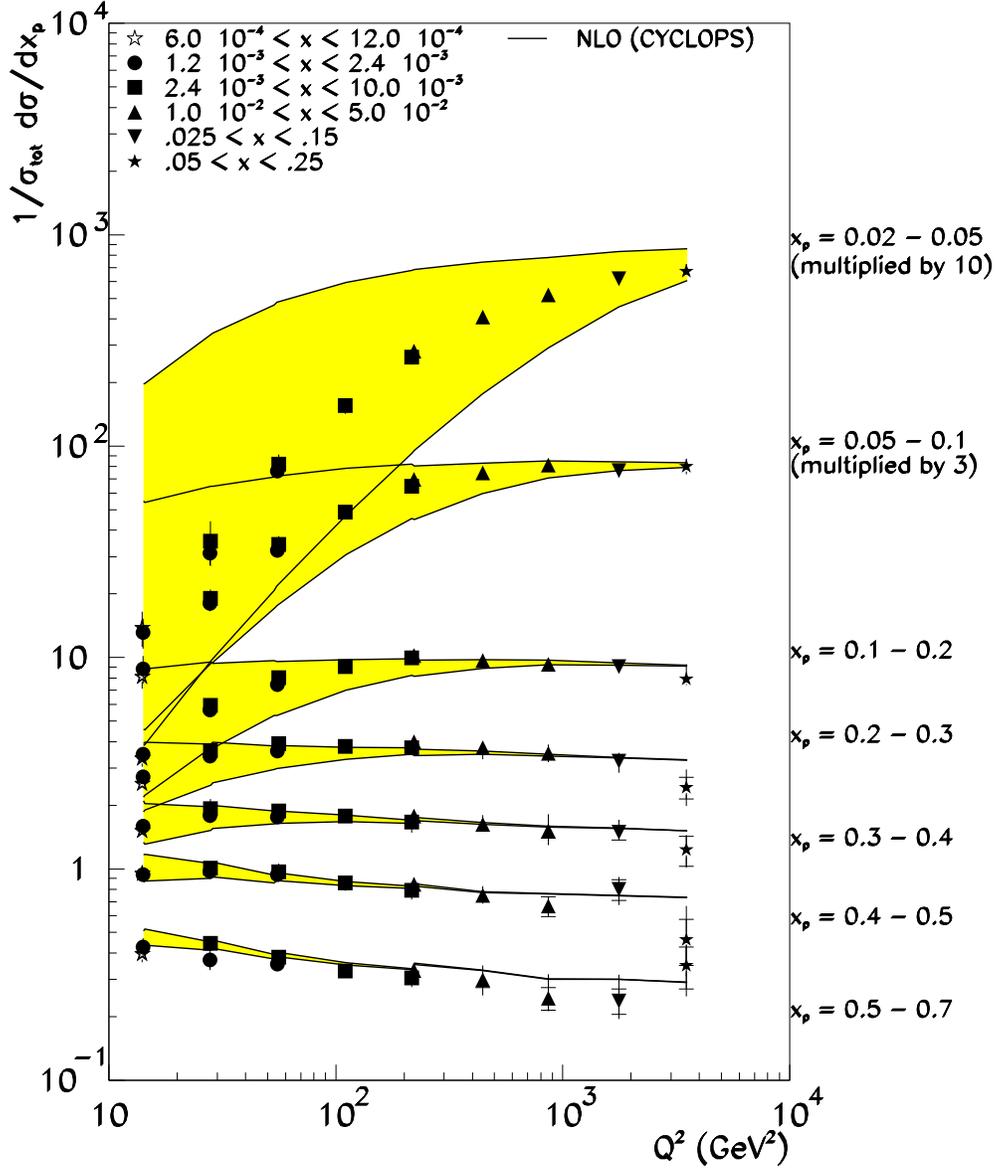,width=0.95\textwidth}}
\caption{The inclusive charged particle distribution,
$ 1/\sigma_{tot}~ d\sigma/dx_p$,
in the current fragmentation region of the Breit frame 
compared to the NLO predictions~\protect\cite{dirk} multiplied by the 
kinematic
correction described in the text. The shaded area represents the extreme cases
$0.1  {\ \rm GeV} < m_{\rm eff} < 1.0  {\ \rm GeV}.$ The upper band
corresponds to $m_{\rm eff} = 0.1 {\ \rm GeV}$
 and the lower band $m_{\rm eff} = 1.0 {\ \rm GeV}.$}
\label{fig:largexp_nlo}
\end{figure}

A kinematic correction has recently been suggested~\cite{durws}
to the NLO calculation~\cite{dirk}  of the inclusive charged particle
distribution which has the form:
\begin{equation}
\frac{1}{1+(\frac{m_{\rm eff}}{Qx_p})^2,}
\end{equation}
where $m_{\rm eff}$ is an effective
mass to account for the massless assumption used in the fragmentation
functions.
It is expected to lie in the range
$0.1 {\ \rm GeV} < m_{\rm eff} < 1.0  {\ \rm GeV}.$
The  $x_p$ data are compared to the CYCLOPS NLO QCD 
calculation incorporating this correction in
Fig.~\ref{fig:largexp_nlo}. This calculation convolutes the fragmentation
function of each type of parton with the cross sections 
for their production. 
It combines a full next-to-leading order matrix element
with the
${\rm MRSA^{\prime}}$ parton densities (with $\Lambda_{\rm QCD} =
230{\rm \ MeV})$
and NLO fragmentation functions
derived from fits to $e^+e^-$ data \cite{binnewies}.
The kinematic correction allows a more legitimate theoretical comparison
to lower $Q^2$ and $x_p$ than was possible in our earlier 
publication~\cite{breit2}. The bands represent the uncertainty in
the predictions by taking the extreme cases of $m_{\rm eff}=0.1  {\ \rm GeV}$
and  $m_{\rm eff}=1.0  {\ \rm GeV}.$ These uncertainties are large at low
$Q^2$ and low $x_p,$ becoming smaller as $Q^2$ and $x_p$ increase.
Within these theoretical uncertainties there is good
agreement throughout the selected kinematic range. 
The kinematic correction describes the general trend of the data but
it is not possible to achieve a good $\chi^2$ fit for $m_{\rm eff}$
over the whole $x_p$
and $Q^2$ range.
The uncertainties introduced by these additional processes
restrict to high $Q^2$ and high $x_p$ the kinematic range 
that may be used to extract $\alpha_s$ from the observed scaling violations.



\begin{figure}[ph!]
\begin{center}\mbox{\epsfig{file=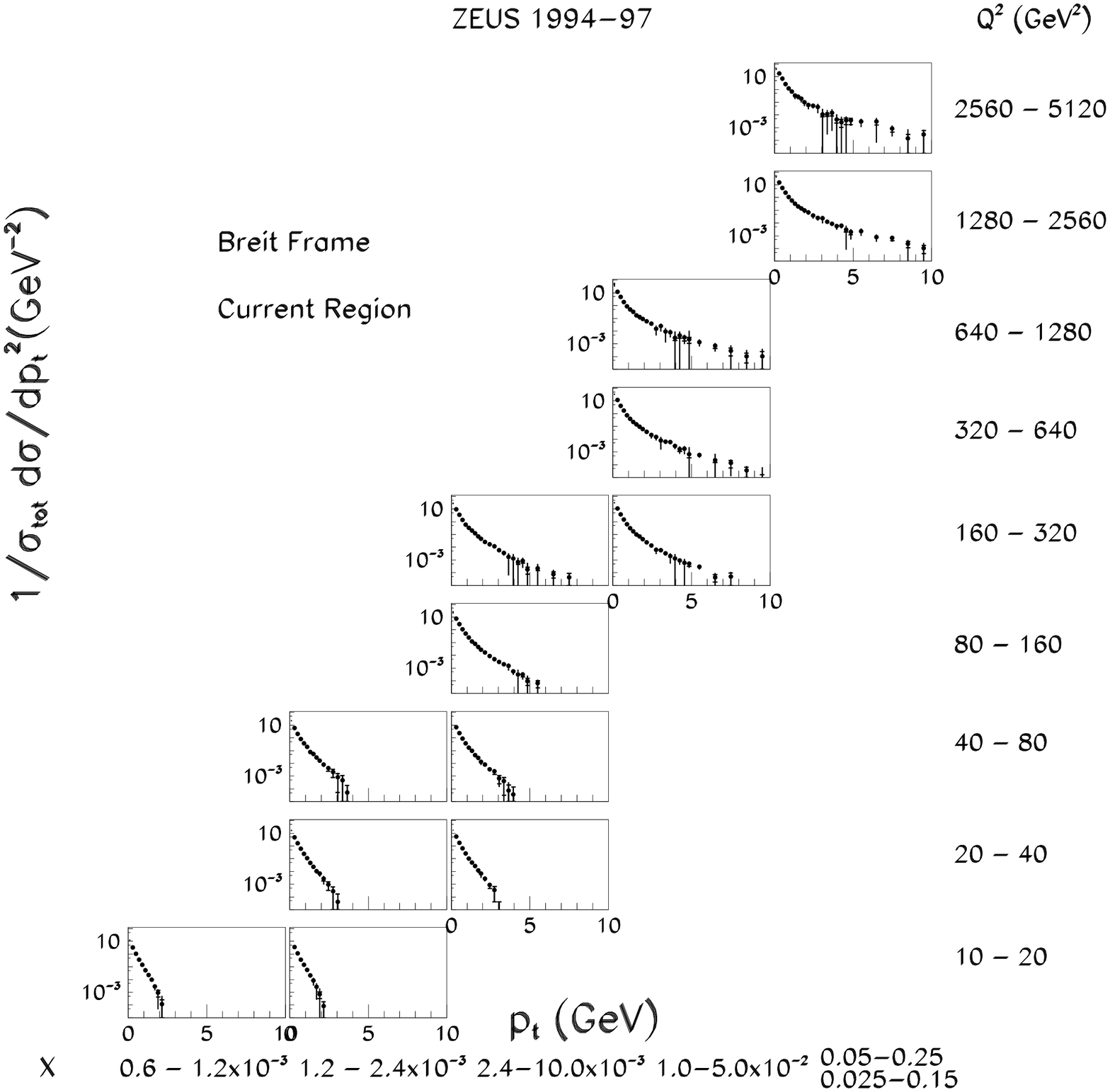,width=14cm,height=14cm}}
\end{center}
\caption{ The transverse momentum distributions in the current
fragmentation region for the 1994-1997 data ($x_p < 1.0$)
for different regions of $x$ and $Q^2.$
The outer error bars are the statistical errors; the inner error bars are
the sum of statistical and
systematic errors added in quadrature.}
\label{ptcurrall}
\end{figure}

The $p_t$ distributions, $1/\sigma_{tot}\ d\sigma/dp_t^2,$ are shown
in Fig.~\ref{ptcurrall} 
 for $x_p < 1.0.$ The distributions show an exponential
fall off at low $p_t$ although it is evident that a high-$p_t$ tail 
develops with increasing $Q^2.$
These high $p_t$ tails contribute at most 15\% of the cross section.
The $p_t$ distributions, from the 1995 data,
in the first 8 bins of $(x,Q^2)$ are shown in Fig.~\ref{ptcurr} 
as closed data points.
The straight lines are exponential fits,  $\exp(-b\sqrt{p_t^2 + m_{\pi}^2}),$
to the low $p_t$ interval 0.2 - 1.0 GeV,
where $m_{\pi}$ is the mass of the pion.
They yield 
slopes of $ b \sim 5-6\ \rm GeV^{-1}.$
The values of $b$
show little $Q^2$ dependence.
For bin 8 the  line extending out to higher $p_t$ ($p_t > \rm 1.2\ GeV$)
is a fit to the empirical power law formula~\cite{photo}
$A\times (1+{p_t}/p_{t0})^{-m}.$
There are strong correlations between $A,p_{t0}\ {\rm and\ }m.$
Consequently $p_{t0}$ has been fixed at 0.75 GeV, a value consistent with 
the fit with all variables free and that used by H1 in the analysis
of their photoproduction data~\cite{H1photo}. 
With this  $p_{t0},$  the parameterisation fits
the data well ($\chi^2/{\rm NdF} = 5.1/19$) and gives
 $m=5.8 \pm 0.4 \pm 0.1.$
These tails are slightly higher if the $x_p < 1.0$ cut is removed
(open points in the figure).
Particles with $x_p > 1.0 $ occur due to hard QCD processes,
such as BGF and QCD Compton.

\begin{centering}
\begin{table}[h!t]
\begin{center}
\begin{tabular}{|c|c|c|}
\hline
  bin no. & $ <Q^2>$ & $<n> \pm $ stat $\pm $ syst\\
 & (GeV$^2$)    &       \\
\hline
\hline

1 & 14.0  &  1.13 $\pm 0.01 \pm 0.05 $ \\
2 & 14.1  &  1.18 $\pm 0.01 \pm 0.04 $ \\
3 & 27.9  &  1.70 $\pm 0.01 \pm 0.07 $ \\
4 & 55.3  &  2.27 $\pm 0.01 \pm 0.07 $ \\
5 & 28.0  &  1.81 $\pm 0.01 \pm 0.06 $\\
6 & 55.9  &  2.44 $\pm 0.01 \pm 0.14 $ \\
7 & 110.  &  3.00 $\pm 0.01 \pm 0.23 $ \\
8 & 216.  &  3.77 $\pm 0.02 \pm 0.26 $ \\
9 & 221.  &  3.98 $\pm 0.02 \pm 0.37 $ \\
10 & 443.  &  4.59 $\pm 0.03 \pm 0.40 $ \\
11 & 863.  &  5.26 $\pm 0.05 \pm 0.39 $\\
12 & 1766. &  6.01 $\pm 0.05 \pm 0.46 $\\
13 & 3507. &  7.10 $\pm 0.11 \pm 0.69 $\\
\hline
\end{tabular}
\caption{Mean charged multiplicity in the current fragmentation region.}
\label{tab:ncurr}
\end{center}
\end{table}
\end{centering}

\begin{figure}[ph!]
\begin{center}\mbox{\epsfig{file=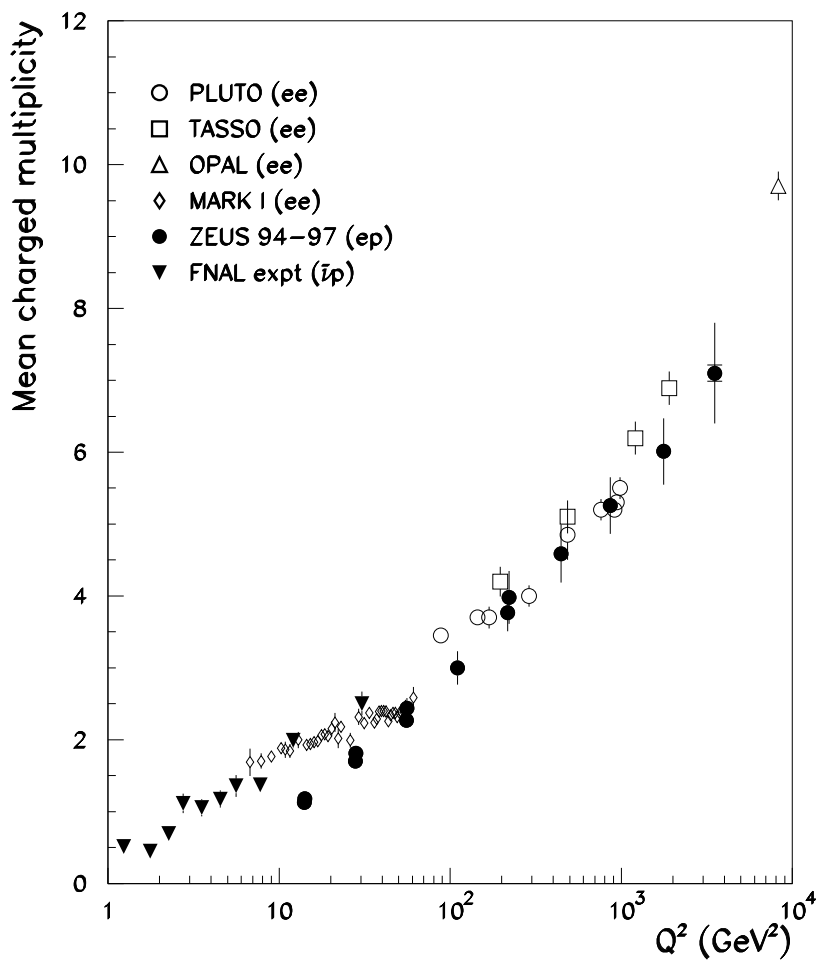,width=14cm}}
\end{center}
\caption{ The mean charged multiplicity for the 
current region. 
The error bars are the sum of the statistical and systematic errors in
quadrature.
The open points represent data from $e^+e^-$ experiments divided by two
to
account for $q$ and $\bar q$ production (also corrected
for contributions from to the charged multiplicity from
$K^0_S$ and $\Lambda$ decays.) Also shown are fixed target DIS
data.}
\label{fig:ncurr}
\end{figure}

Figure~\ref{fig:ncurr} shows the mean charged multiplicity 
in the current fragmentation region.
The results, for each of the $(x,Q^2)$ bins, are listed in
Table~\ref{tab:ncurr}. The multiplicity increases by about a factor of
six over the measured $Q^2$ range.
Also shown in Fig.~\ref{fig:ncurr} are results 
from $e^+e^-$ annihilation experiments~\cite{nchee}  (scaled down
by a factor of 2) and results 
from fixed target DIS data~\cite{musgrave} at similar 
$Q^2 (Q^2 < 30 {\rm\ GeV^2})$ to the ZEUS data
but corresponding to an $x$ range 
about two orders of magnitude higher.
For $Q^2 \sgeq 80{\rm\ GeV^2}$ there is reasonable agreement between the 
results from $e^+e^-$  data and ZEUS, again consistent with the universality
of fragmentation. 
At lower $Q^2$ the multiplicities measured
by ZEUS are lower than those found in the
 $e^+e^-$  data and the fixed target DIS data.
Similar results have recently been observed by the NOMAD
collaboration~\cite{NOMAD}.
In this $Q^2$ region
there is a negligible contribution from charmed quarks so that the
difference must originate from the depopulation of the current region
due to the prevalence of boson-gluon fusion processes in this low $(x,Q^2)$ 
region~\cite{review,val,eden}.
  
\begin{figure}[ph!]
\begin{center}\mbox{\epsfig{file=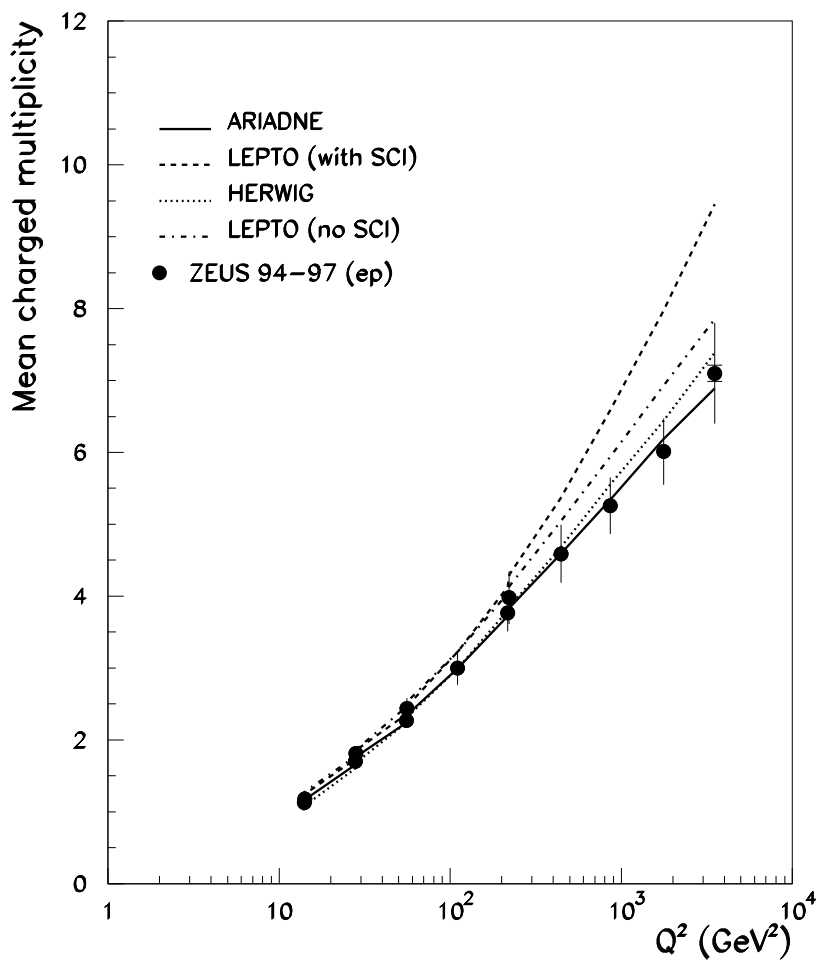,width=14cm}}
\end{center}
\caption{ The mean charged multiplicity for the 
current region. 
The error bars are the sum of the statistical and systematic errors in
quadrature. Also shown are the predictions from 
the Monte Carlo generators ARIADNE (full line), LEPTO with SCI (dashed lines),
HERWIG (dotted line) and LEPTO with no SCI (dash-dotted line). }
\label{fig:nchmc}
\end{figure}

Figure~\ref{fig:nchmc} displays the same ZEUS data as in Fig.~\ref{fig:ncurr}
but now
compared to various Monte Carlo models. Both ARIADNE and
HERWIG, with default settings, describe well the variation of the multiplicity
with $Q^2.$ LEPTO with SCI, 
while describing the data at low $Q^2,$ simulates the $Q^2$
evolution incorrectly which
leads to an overestimation of the multiplicity at
high $Q^2.$ This overestimation of the data by LEPTO can be partially
rectified, as
can be seen in Fig.~\ref{fig:nchmc}(dash-dotted line), 
by removing SCI from the model.

\subsection{Target fragmentation region}

\begin{figure}[b]
\begin{center}\mbox{\epsfig{file=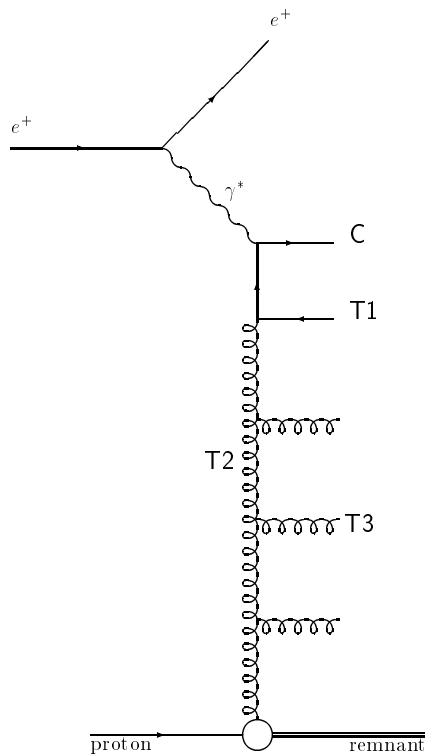,height=0.45\textheight}}
\end{center}
\caption{{ A schematic of
DIS scattering at low $x$ within the MLLA framework.
Quark C represents the struck sea 
quark in the current fragmentation region. T1 is
the other half of the quark box which is in the
target region. T2 is the t-channel
gluon exchange and T3 the rungs of the gluon ladder.}}
\label{pic1}
\end{figure}

\begin{figure}[ph!]
\begin{center}\mbox{\epsfig{file=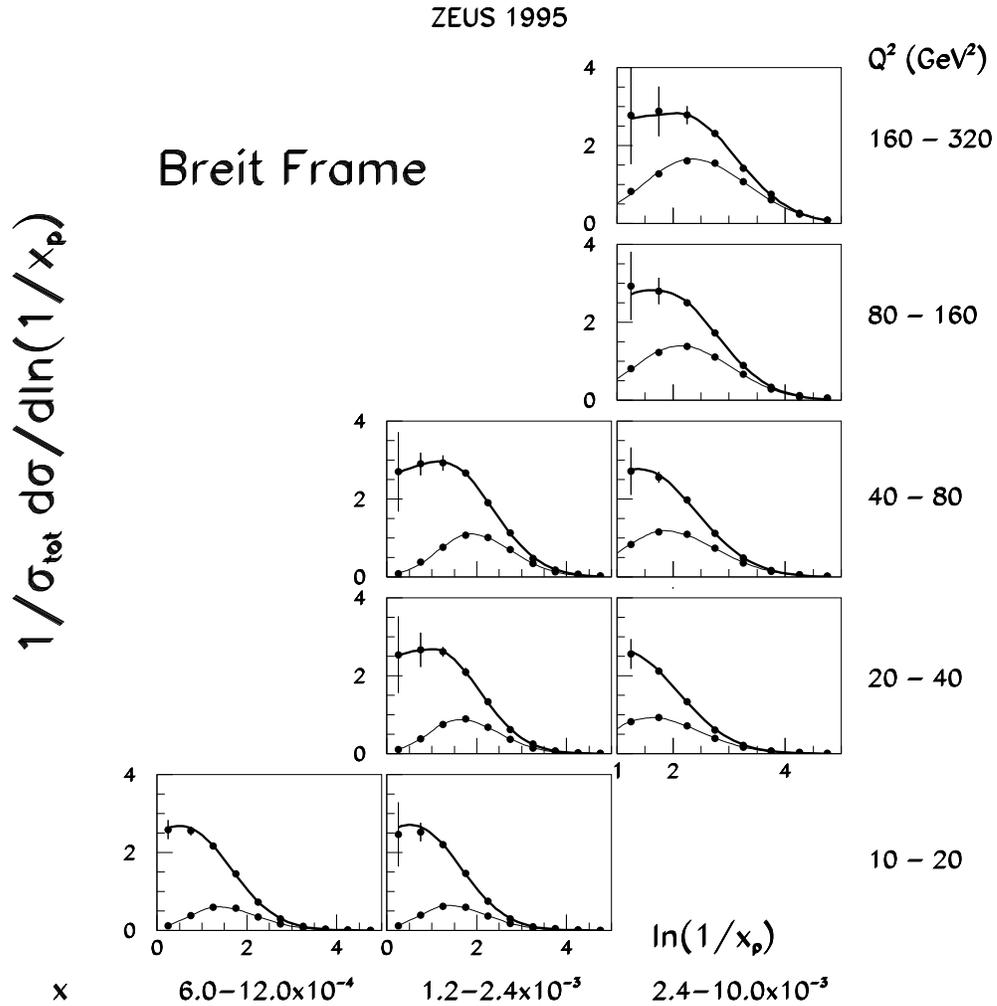,width=14cm,height=14cm}}
\end{center}
\caption{ The corrected $\ln(1/x_p)$ distributions  for the target
and current
regions for the 1995 data. Fitted two-piece normal distributions are shown
to guide the eye. The heavy line
corresponds to
the target region, the light line to the current region.
The error bars are the sum of the statistical and systematic errors in
quadrature.}
\label{lxpda}
\end{figure}

\begin{figure}[ph!]
\begin{center}\mbox{\epsfig{file=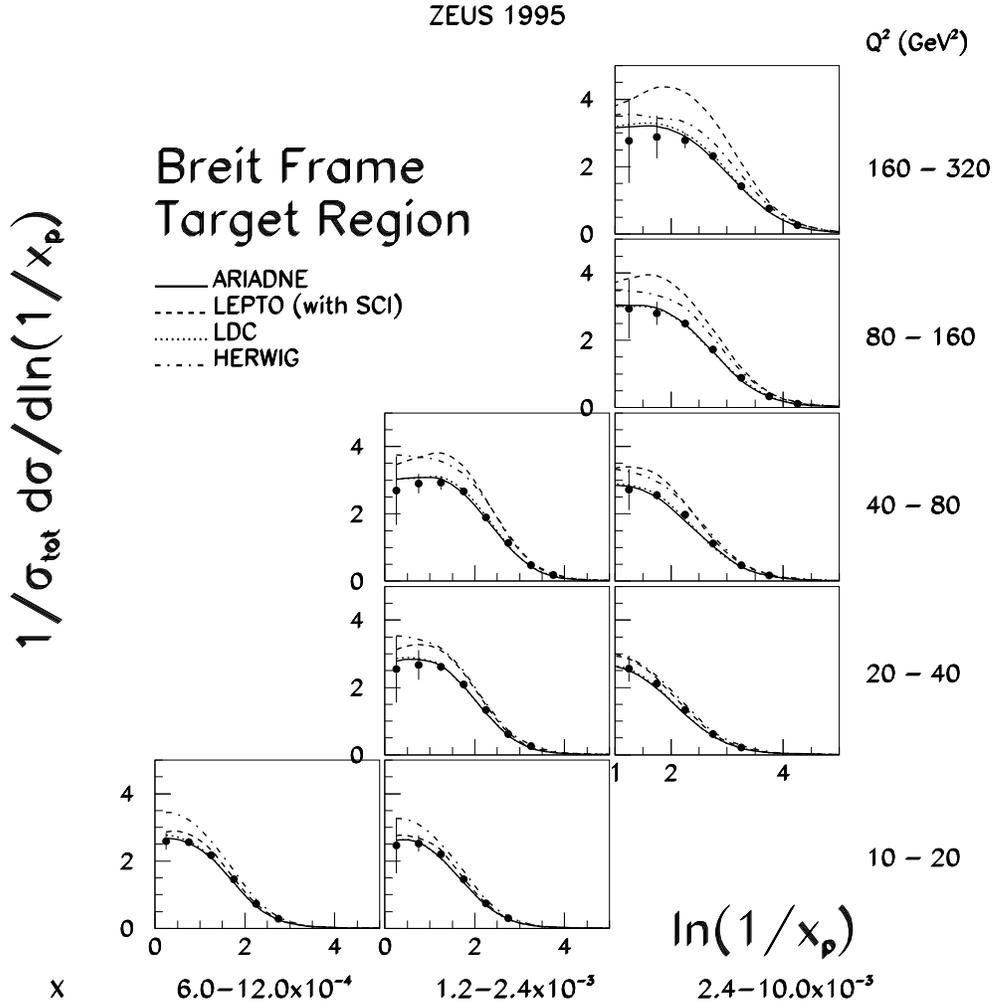,width=14cm,height=14cm}}
\end{center}
\caption{ The corrected $\ln(1/x_p)$ distributions  for the target
fragmentation
region for the 1995 data compared to Monte Carlo models:
ARIADNE, LEPTO with SCI, LDC and HERWIG.
The error bars are the sum of the statistical and systematic errors in
quadrature.
The LEPTO model without SCI resembles the predictions
of ARIADNE.}
\label{fig:logxp_mc}
\end{figure}

\begin{figure}[ph!]
\begin{center}\mbox{\epsfig{file=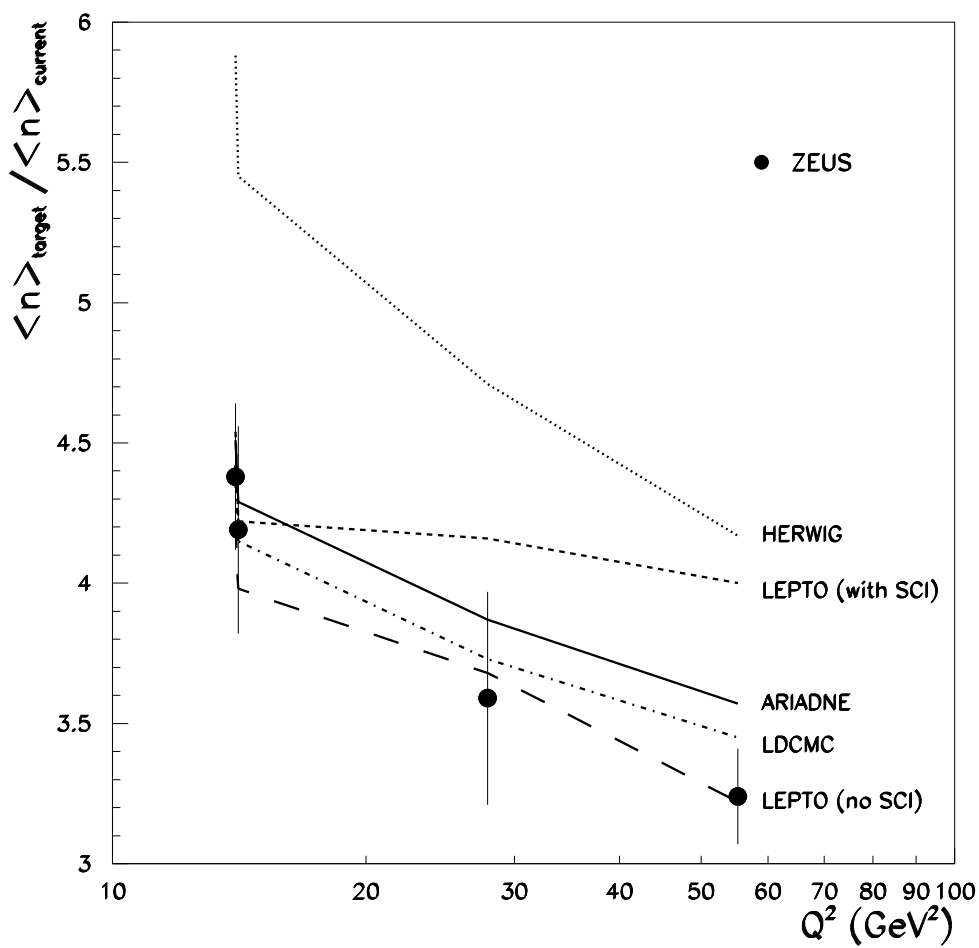,
width=14cm,height=14cm}}
\end{center}
\caption{ The ratio of the charged multiplicities in 
the target ($x_p<1$)
and current
regions of the Breit frame as a function of $Q^2.$ The
data are compared to Monte Carlo models:
ARIADNE, LEPTO with and without SCI, LDC and HERWIG.
The error bars are the sum of the statistical and systematic errors in
quadrature. The discontinuities, at the lowest $Q^2,$
 in the Monte Carlo curves are due to 
overlapping $Q^2$ bins at different values of $x.$}
\label{fig:ratnch}
\end{figure}

\begin{figure}[ph!]
\begin{center}\mbox{\epsfig{file=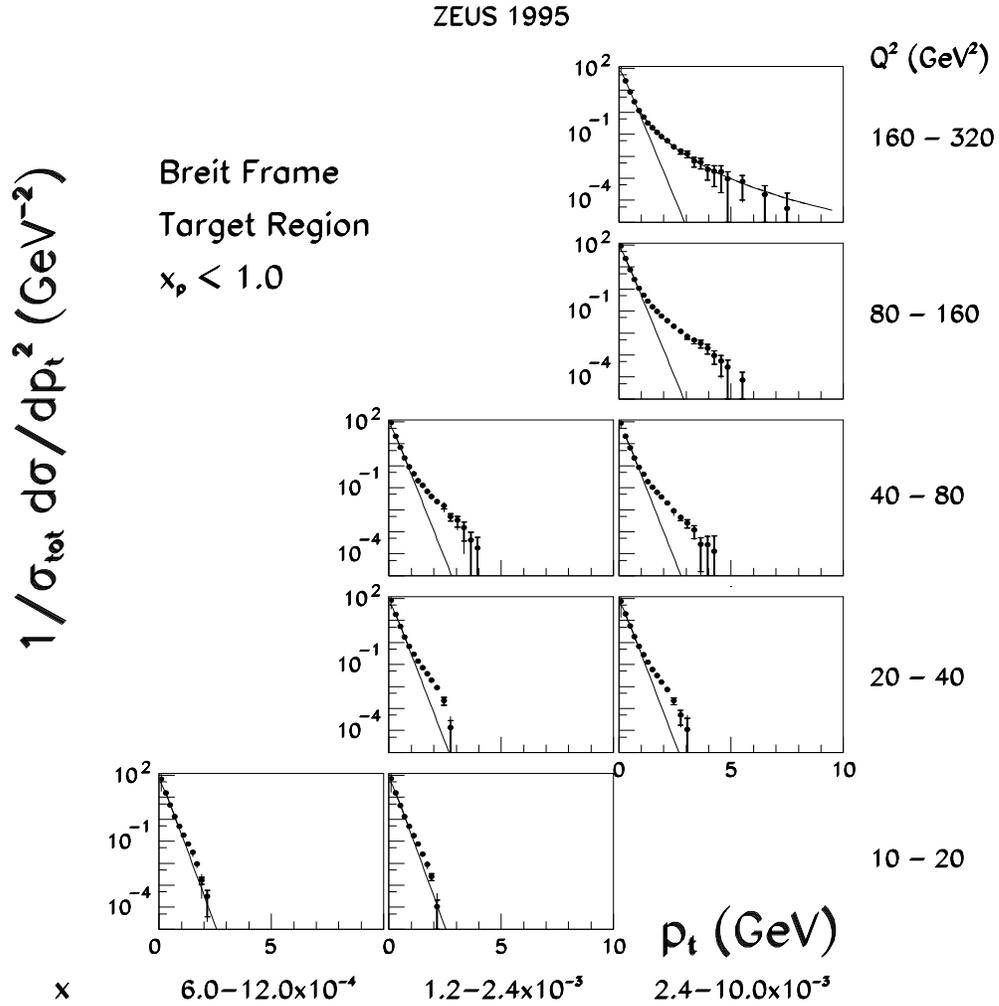,width=14cm,height=14cm}}
\end{center}
\caption{ The transverse momentum distributions in the target
fragmentation region for the 1995 data.
The inner error bars are the statistical errors; the outer error bars are
the sum of statistical and
systematic errors added in quadrature. The lines are the fits discussed in
the text.}
\label{pt2da}
\end{figure}

\begin{figure}[ph!]
\begin{center}\mbox{\epsfig{file=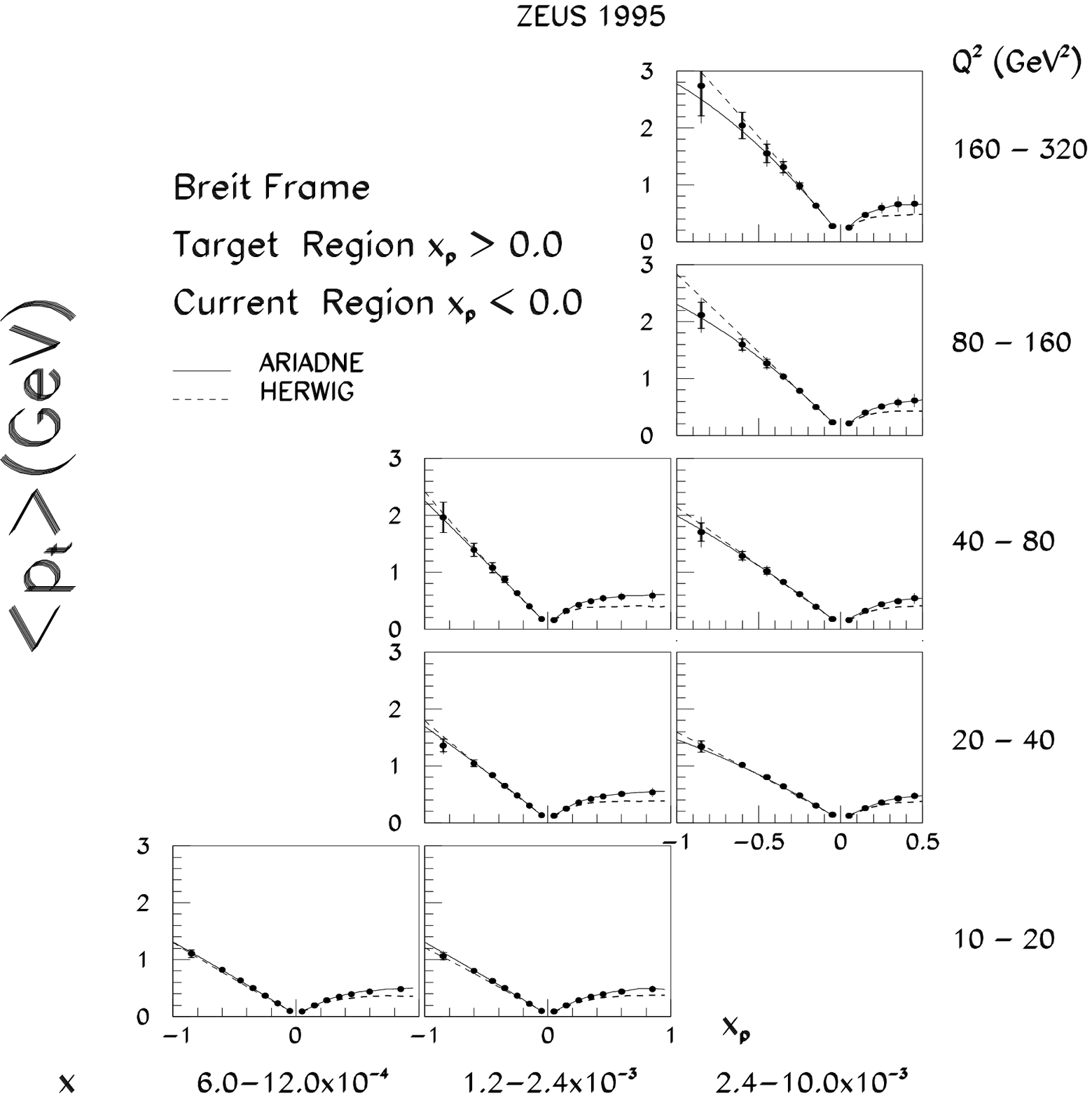,width=14cm,height=14cm}}
\end{center}
\caption{ The corrected  mean transverse momentum  versus  scaled
momentum
distributions for the 1995 data.
The inner error bars are the statistical errors; the outer error bars are
the sum
 of statistical and
systematic errors added in quadrature. The full line indicates the
ARIADNE Monte Carlo prediction and the dotted the HERWIG prediction.}
\label{mptda}
\end{figure}

The distributions of charged particles in $x_p$
and transverse momentum, $p_t$, in the target
region of the Breit frame are  studied as a function of $x$ and $Q^2$,
in the range
$6\times10^{-4} < x < 1\times10^{-2}$ and \mbox{$10<Q^2<320~$}
GeV$^2$.
This analysis, which uses the 1995 data, is restricted to
values of the scaled momenta $x_p  < 1.0$ so that 
similar phase space regions for 
the target and current can be compared.
Thus the accepted part of the target region
corresponds to the contributions T1 and T2 in Fig.~\ref{pic1}
and is only a small part of the complete phase space, as depicted
in Fig.~\ref{pic0}.
This is also the
kinematic region of the target fragmentation
that has a reasonable acceptance. 
The corrected data distributions with combined statistical
and systematic errors are shown in Figs.~\mbox{\ref{lxpda}-\ref{mptda}.}

The distributions in $\ln(1/x_p)$
are shown for both 
the target and current regions in Fig.~\ref{lxpda}. The
fitted curves shown are
two-piece normal distributions~\cite{boe} to guide the eye. 
In contrast to the current region,
the target region distribution does not fall
to zero as $\ln(1/x_p)$ tends to zero.
Although the magnitude of the single particle density at the
peak position of the current region distribution
grows by a factor
of about three over the $Q^2$ range shown, the single particle
density  of the target
distribution, at the $x_p$ value corresponding to the peak of the
current distribution
(contribution C is equivalent to contribution T1 in Fig.~\ref{pic1}),
depends less strongly on $Q^2$ and 
increases by only about $30\%$.
In addition the $\ln(1/x_p)$ distribution 
shows no significant dependence on $x$ when $Q^2$ is kept constant.
In the target region the peak position of the $\ln(1/x_p)$ distribution
increases more rapidly with $Q^2$ than in the current region; this is
consistent with the behaviour expected from cylindrical phase space.
The approximate Gaussian distribution of the MLLA predictions peaking
at $\ln(1/x_p) \sim 1.5-2.5$~\cite{anis} is not observed. We
conclude that the target
distributions are inconsistent with the MLLA predictions  when used
in conjunction with LPHD.

The $\ln(1/x_p)$ distributions in the
target fragmentation region are compared
to Monte Carlo models in Fig.~\ref{fig:logxp_mc}. The ARIADNE and
Linked Dipole Chain (LDC) models describe the data well in the 
measured $(x,Q^2)$ bins. 
The two Monte Carlo models based on DGLAP parton evolution
techniques, LEPTO and HERWIG, fail to describe the data. 
The LEPTO Monte Carlo with SCI
describes the data at low $Q^2$ but
the dependence on $Q^2$ within the model is incorrect and discrepancies
are observed at large $Q^2.$ The HERWIG Monte Carlo gives a poor description
of the data in all $(x,Q^2)$ bins.
The LEPTO generator without SCI (not shown) gives a good
description of the data.

\begin{table}[t!]
\begin{center}
\begin{tabular}{|c|c|c|c|}
\hline
 bin no. & $ <Q^2> $ (GeV$^2$) & current $<n>$&
target $<n>$\\
\hline\hline
 1 & $ 14.0 $ & 1.13 $\pm$0.01$\pm$0.05  &
  4.95 $\pm$0.01$\pm$0.2  \\
\hline
 2 & $ 14.1 $ & 1.18 $\pm$0.01$\pm$0.04  &
4.94 $\pm $0.01$\pm$0.4 \\
 3 & $ 27.9 $ &1.70 $\pm$0.01$\pm$0.07  & 6.11 $\pm$0.02$\pm$0.6
\\
 4 & $ 55.3 $ &  2.27 $\pm$0.01$\pm$0.07 & 7.36 $\pm$0.03$\pm$0.3
\\
\hline\hline
\end{tabular}
\caption { The mean charged multiplicities in the current and
target regions for the
($x$,$Q^2$) analysis bins, in the range $0 <x_p <1.$
The first error is statistical and the second is systematic.}
\label{tab:ntrgt}
\end{center}
\end{table}

The mean multiplicity 
in the target region in the range $ 0 <x_p <1,$  shown in
Table~\ref{tab:ntrgt}, is larger
than in the current region, by about a factor of four in the lowest
$(x,Q^2)$ bins.
The target region multiplicity increases with $Q^2$, but more slowly than that
in the current region presumably due to the additional fragmentation terms in 
the target region shown in Fig.~\ref{pic1}.
Only the first four $(x,Q^2)$ bins are studied as they
have reasonable acceptance over the whole of $ 0 <x_p <1.$
Fig.~\ref{fig:ratnch} shows the ratio of the charged multiplicities
in the target and the current regions as a function of $Q^2.$
The ratio falls as $Q^2$ increases. Also shown are comparisons with Monte Carlo
models; ARIADNE, LDC and LEPTO without SCI 
all describe the trend of the data. 
HERWIG, though reproducing the $Q^2$ dependence, fails to predict 
the magnitude whilst
LEPTO with SCI
fails to describe the $Q^2$ dependence of the ratio.

\begin{figure}[ph!]
\begin{center}\mbox{\epsfig{file=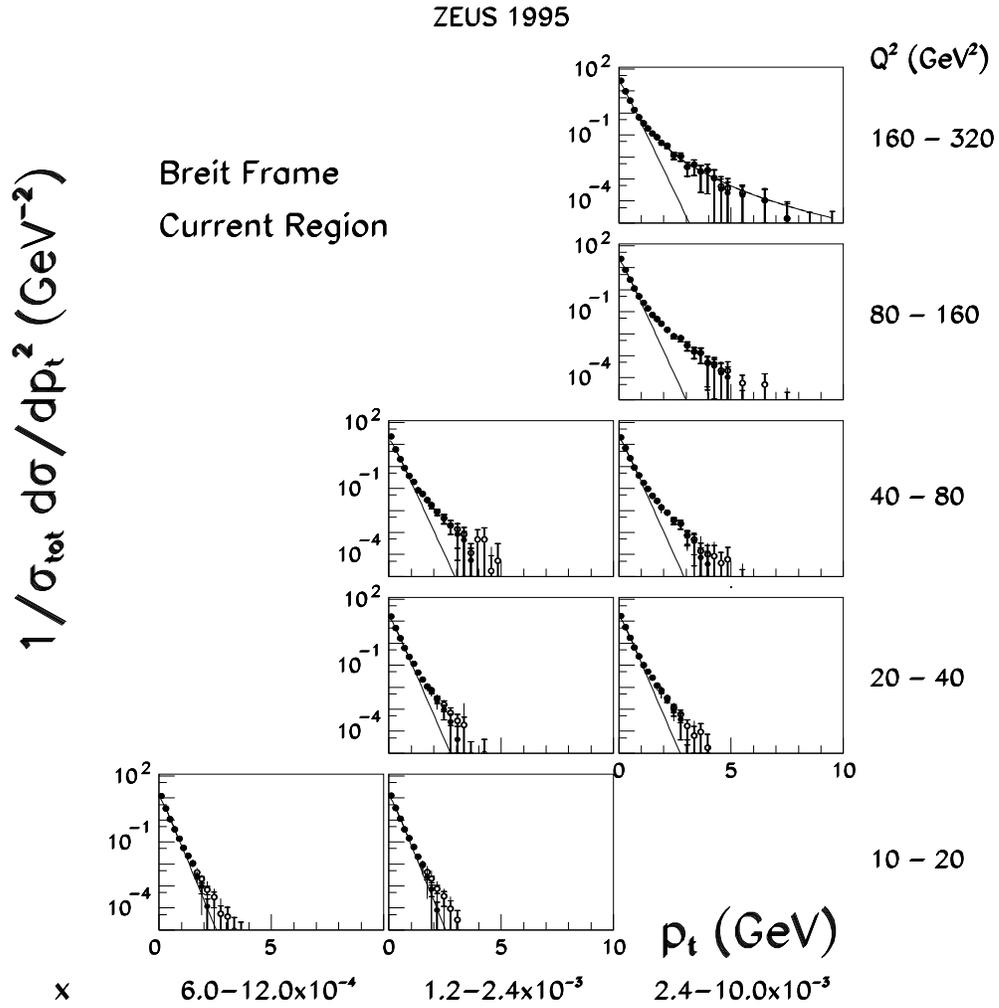,width=14cm,height=14cm}}
\end{center}
\caption{ The transverse momentum distributions in the current
fragmentation region for the 1995 data for different regions
of $Q^2$ and $x$.
The inner error bars are the statistical errors; the outer error bars are
the sum of statistical and
systematic errors added in quadrature. The lines are the fits discussed in
the text. The closed data points are for tracks with $x_p < 1.0$ and
the open data points are for all $x_p.$}
\label{ptcurr}
\end{figure}

The $p_t$ distributions,
$1/\sigma_{tot} d\sigma/dp_t^2$, are shown in Fig.~\ref{pt2da} 
and the same fits
have been performed on these distributions as on the 
current region $p_t$ distributions of Fig.~\ref{ptcurr}.
The fit of the exponential,  $\exp(-b\sqrt{p_t^2 + m_{\pi}^2}),$
gives slopes of $ b \sim 5-6 \rm\ GeV^{-1}.$
In a similar manner to the distributions in the current region,
the values of $b$
exhibit little $Q^2$ dependence,
but the distributions develop a high-$p_t$ tail with increasing $Q^2$.
The  line
plotted for the $Q^2$ interval 160-320 $\rm GeV^2$ is a fit to the power law
formula $A\times (1+{p_t}/p_{t0})^{-m}$ for $p_t$ greater than 1.2 GeV.
This fits the data well 
($\chi^2/{\rm NdF} = 1.9/19$) with
$p_{t0}$ fixed at 0.75 GeV and $ m = 5.7 \pm 0.6\pm 0.1.$
These high $p_t$ tails contain at most 15\% of the cross section.
The values of $m$ in the current and target regions agree within errors
 and are smaller than that found for the ZEUS~\cite{photo} and
H1~\cite{H1photo} photoproduction data.
This is consistent with what would be expected from the point-like
nature of the exchanged photon in DIS.

To compare the general characteristics of the transverse momentum
distributions in the
target and current regions, the mean $p_t$ versus $x_p$ is shown in 
Fig.~\ref{mptda}.
The mean $p_t$ at large $x_p$ is
higher in the
current region than in the target region and shows a stronger $Q^2$
dependence than
the target region. In the current region the mean $p_t$ rises with $x_p$
and reaches a
maximum as $x_p$ tends to $-1.$ In contrast, in the target region the mean
$p_t$ tends to
a constant value of about 0.6 GeV. Thus the target region $p_t$
distribution has the
characteristics of $p_t$-limited phase space with only a small
dependence on $Q^2$.
The mean $p_t$ vs $x_p$ distribution in the target region
shows no significant dependence on $x$ when
$Q^2$ is kept constant.

Also shown in Fig.~\ref{mptda} are the Monte Carlo predictions of the
ARIADNE and HERWIG models. The ARIADNE generator gives a good
description of the data and is very similar to the predictions of LEPTO
and LDC Monte Carlo models. The HERWIG generator gives a less
satisfactory description of the data. The discrepancy in the target
region can partially be explained by the lack of intrinsic transverse
momentum of the incoming struck quark in the default parameters of
HERWIG.

The apparent contradiction between the similarity of Figs.~\ref{ptcurr}
and~\ref{pt2da} and the difference between the mean $p_t$ in the current
and target regions in Fig.~\ref{mptda} may be understood in terms of
the correlation between $x_p$ and $p_t.$
In the current region this correlation is strong
with the high $x_p,p_t$ region corresponding to low multiplicity.
As the
high $p_t$ particles have a strong $Q^2$ dependence, this is reflected
in the $Q^2$ dependence of the mean $p_t$ at high $x_p.$ In contrast, in
the target region the correlation between $x_p$ and $p_t$ is small.
This results in a lower mean $p_t$ at large $x_p$ and a mean $p_t$
substantially independent of $x_p$ and $Q^2$ as $x_p$ tends to 1.
  
\section{Summary}
Charged particle distributions have been studied in the 
Breit frame in DIS over a wide range of $Q^2.$ 
The distributions in  scaled momentum,
$x_p$,  and
transverse momentum, $p_t$,  have been measured for the first time
in the target region of the Breit frame for $1.2\times10^{-3} < x <
1\times 10^{-2}$ and $10 < Q^2 < 320 {\rm\ GeV^2}.$
For scaled momenta in the interval $0 < x_p <1$ the mean
target region charged track
multiplicity is found to be larger than that measured in the
current region; there is no significant $x$-dependence at
fixed $Q^2.$

The transverse momentum distributions for both the current and
target fragmentation regions exhibit similar properties.
A tail at large $p_t$ develops as $Q^2$ increases.
The mean transverse momentum as a function of $x_p$
has a weaker dependence on $Q^2$ in the target region
than the current region. 
Whereas  in the current region the mean
$p_t$ increases
approximately linearly with $x_p$, the mean $p_t$ in the target region tends to a constant
value with increasing $x_p,$ consistent with cylindrical phase space.

The HERWIG model is unable to describe the $Q^2$
dependence of the $\ln(1/x_p)$ distributions in the target fragmentation
region. 
In contrast the colour dipole model as implemented in the ARIADNE program,
LEPTO generator without SCI
and the LDC Monte Carlo, based on a reformulation of the CCFM
evolution, all adequately describe the data.

In the current region, the results show clear
evidence for scaling violations in scaled momenta as a function of $Q^2$
and support the hypothesis of the coherent nature of QCD cascades. 
The data are well described by NLO calculations.
The comparison of our results in the
current region of the Breit frame 
with  $e^+e^-$ data at $Q^2 = s_{ee}$
for $Q^2 > 80{\rm \ GeV^2}$  
shows good agreement.
The moments of the $\ln(1/x_p)$ spectra in the current region
exhibit the same energy scale behaviour as those observed
in $e^+e^-$ data. 
The observed charged particle spectra are consistent
with the universality of quark fragmentation in $e^+e^-$ and DIS
at high $Q^2.$
The moments cannot be described by 
the MLLA calculations together with LPHD.

The  target region $\ln(1/x_p)$ distribution shows a weaker $Q^2$ dependence
than  the
corresponding current region distribution.
In particular, the magnitude of the single particle density of the target
distribution, at the $x_p$ value corresponding to the peak of the
current
distribution,  increases by  about $30\%,$
in contrast to  a threefold increase for the current region in the
$Q^2$-range considered here.
The MLLA predictions for the target region, in conjunction with LPHD,
fail to describe the data.

\section*{Acknowledgements}
The strong support and encouragement of the DESY Directorate have
been invaluable, and we are much indebted to the HERA machine group
for their inventiveness and diligent efforts.  The design,
construction and installation of the ZEUS detector have been made
possible by the ingenuity and dedicated efforts of many people from
inside DESY and from the home institutes who are not listed as authors.
Their contributions are acknowledged with great appreciation.
 
This paper was completed shortly after the tragic and untimely death of
Prof.~Dr. B.~H.~Wiik, Chairman of the DESY directorate. All members of
the ZEUS collaboration wish to acknowledge the remarkable r\^ole which
he played in the success of both the HERA project and of the ZEUS
experiment. His inspired scientific leadership, his warm personality and
his friendship will be sorely missed by us all.

\newpage

\newpage 

\newpage

\newpage

\newpage

\newpage

\newpage

\newpage

\newpage

\newpage

\newpage

\newpage

\newpage

\newpage

\newpage


\begin{thebibliography}{99}
\bibitem{feyn} R.~P. Feynman, `Photon-Hadron Interactions', Benjamin,
N.Y.  (1972).

\bibitem{breit1} ZEUS Collab., M.~Derrick  et al.,  Z.\ Phys.\ C67 (1995) 93.

\bibitem{breit2} ZEUS Collab., M.~Derrick  et al.,  Phys.\ Lett.\ B414 
(1997) 428.

\bibitem{H1breit}  H1 Collab., S.~Aid et al., Nucl. Phys. B445 (1995) 3;\\
H1 Collab., C.~Adloff et al., Nucl. Phys. B504 (1997) 3.

\bibitem{eedis} Yu.~Dokshitzer  et al.,  Rev.\ Mod.\ Phys.\ 60 (1988) 373.

\bibitem{anis} A. V. Anisovich  et al.,  Il Nuovo Cimento A106 (1993) 547.

\bibitem{char} K.\ Charchu{\l}a,  J.\ Phys.\  G19 (1993) 1587.

\bibitem{val}  K.H. Streng, T.F. Walsh and P.M. Zerwas, 
Z. Phys. C2~(1979)~237.

\bibitem{bassetto} A. Bassetto et al., Nucl. Phys. B207 (1982) 189.

\bibitem{mueller} A. Mueller, Nucl. Phys. B213 (1983) 85.

\bibitem{webber} B.~R.~Webber, Nucl. Phys. B238 (1984) 492.

\bibitem{MLLA} Yu.~Dokshitzer, V. Khoze, A. Mueller and S. Troyan,
``Basics of Perturbative QCD'', Editions Fronti\`{e}res, Gif-sur-Yvette,
France (1991).

\bibitem{review} V. Khoze and W. Ochs, Int. J. Mod. Phys.  A12 (1997) 2949.

\bibitem{fongweb} C.\ P.\ Fong and B.~R.~Webber, Phys. Lett. B229 (1989) 289;\\
C.\ P.\ Fong and B.~R.~Webber, Nucl. Phys. B355 (1991) 54.

\bibitem{dokevol} Yu. Dokshitzer, V.~Khoze and S.~Troyan, Int. J. Mod. Phys. A7 (1992) 1875.
 
\bibitem{LPHD} Ya.~Azimov et al., Z. Phys. C27 (1985) 65.

\bibitem{dok} Yu. Dokshitzer  et al., Sov. Phys.\ JETP 68 (1988) 1303.

\bibitem{DGLAP}
V.N.~Gribov and L.N.~Lipatov,
\newblock  Sov.\ J.\ Nucl.\ Phys.\ 15 (1972) 438 and 675;\\
\newblock Yu.L.~Dokshitzer, Sov.\ Phys.\ JETP\  46 (1977) 641;\\
G.~Altarelli and G.~Parisi,  Nucl.\ Phys.\  B126 (1977) 298.

\bibitem{webnas} G.~Altarelli et al., Nucl. Phys. B160 (1979) 301;  \newline
P.~Nason and B.~R.~Webber, Nucl. Phys. { B421}~(1994) 473.

\bibitem{ALEPH} ALEPH Collab., D.~Buskulic et al., Phys. Lett.
{ B357}~(1995)~487.

\bibitem{DELPHI} DELPHI Collab., P.~Abreu et al., Phys. Lett. { B311}
(1993) 408;
\newline DELPHI Collab.,   P.~Abreu et al., Phys. Lett. B398 (1997) 194.

\bibitem{dirk} D.~Graudenz, CERN--TH/96--52;
\newline D.~Graudenz, CYCLOPS program and private communication.

\bibitem{b:sigtot_photoprod} ZEUS Collab., M. Derrick  et al.,
                    Phys.\ Lett.\  B293 (1992) 465;\\
 ZEUS Collab., M. Derrick  et al., Z.\ Phys.\  C63 (1994) 391.

\bibitem{b:Detector} ZEUS Collab., The ZEUS Detector,
                     Status Report 1993, DESY 1993.

\bibitem{b:CTD} N. Harnew et al., Nucl. Inst. Meth. A279 (1989) 290;\\
                B. Foster et al., Nucl.~Phys.~B (Proc.~Suppl.) 32 (1993) 181;\\
                B. Foster et al., Nucl.~Inst.~Meth. A338 (1994) 254.

\bibitem{b:CAL} M. Derrick et al., Nucl. Inst. Meth. A309 (1991) 77;\\
                A. Andresen et al., Nucl. Inst. Meth. A309 (1991) 101;\\
                A. Bernstein et al., Nucl. Inst. Meth. A336 (1993) 23.

\bibitem{jb} F. Jacquet and A. Blondel, Proceedings of the study
for an ep facility for Europe, DESY~79/48~(1979)~391.

\bibitem{DA} S.~Bentvelsen, J.~Engelen and P.~Kooijman,
Proceedings of the 1991 Workshop on Physics at HERA, DESY
Vol.~1~(1992)~23.

\bibitem{z_shift}
ZEUS Collab., M.\ Derrick  et al., Z.\ Phys.\  C72 (1996) 399.

\bibitem{GEANT} R. Brun et al., GEANT3, CERN DD/EE/84-1~(1987).

\bibitem{DJANGO} K.\ Charchu{\l}a, G.\ Schuler and H.\ Spiesberger,
                   Comp.\ Phys.\ Comm.\ 81 (1994) 381.

\bibitem{HERACLES} A.~Kwiatkowski, H.~Spiesberger and H.-J.~M\"{o}hring,
                 Proceedings of the 1991 Workshop on Physics at HERA,
                   DESY Vol.~3~(1992)~1294; \\
A.~Kwiatkowski, H.~Spiesberger and H.-J.~M\"{o}hring, Z. Phys. C50 (1991) 165.

\bibitem{ariadne} L.L\"{o}nnblad, ARIADNE version 4.03 program and manual,
{ Comp. Phys. Comm.} { 71} (1992) 15.


\bibitem{string} B.~Andersson et al., Phys. Rep. 97 (1983) 31.

\bibitem{JETSET}  T.~Sj\"{o}strand, Comp. Phys. Comm. 82 (1994) 74;
\newline T.~Sj\"{o}strand, CERN-TH 7112/93 (revised August 1995).

\bibitem{grv94} M.~Gluck, E.~Reya and A.~Vogt, Z. Phys. C67 (1995) 433.

\bibitem{mrsa}  A.~D.~Martin, R.~G.~Roberts and W.~J.~Stirling,
Phys. Rev. { D50} (1994) 6734.

\bibitem{herwig} G.~Marchesini et al., Comp. Phys. Comm. 67 (1992) 465.

\bibitem{cluster} G.~Marchesini and B.~R.~Webber, Nucl. Phys. B310 (1988) 461.

\bibitem{f2} ZEUS Collab., M. Derrick et al., Z. Phys. C65 (1995) 379.

\bibitem{h1f2} H1~Collab., I.~Abt~et~al., Nucl.~Phys.~B407~(1993)~515.

\bibitem{LEPTO}
\newblock G.~Ingelman, A.~Edin and J.~Rathsman, LEPTO 6.5\_1,
Comp.\ Phys.\ Comm.\ 101 (1997) 108.

\bibitem{SCI} A.~Edin, G.~Ingelman  and J.~Rathsman, Phys. Lett. B366
(1996) 371.

\bibitem{LDC}
B.~Andersson, G.~Gustafson and J.~Samuelsson,
\newblock  Nucl.\ Phys.\ B463 (1996) 217; \\
\newblock B.~Andersson, G.~Gustafson, H.~Kharraziha and J.~Samuelsson,
  Z.\ Phys.\  C71 (1996) 613; \\ 
H.~Kharraziha and L.~L\"onnblad, JHEP 9803 (1998) 006.

\bibitem{CCFM_LDC}
B.~Andersson, G.~Gustafson and J.~Jannelson,
\newblock Nucl.\ Phys.\ B463 (1996) 217.

\bibitem{CCFM}
M.~Ciafaloni, Nucl.\ Phys.\  B296 (1988) 49;\\
\newblock S.~Catani, F.~Fiorani and G.~Marchesini, 
Phys.\ Lett.\  B234 (1990) 339
  and Nucl.\ Phys.\  B336 (1990) 18; \\ 
G.~Marchesini, Nucl.\ Phys.\  B445 (1995) 49.

\bibitem{BFKL}
{E.A.~Kuraev, L.N.~Lipatov and V.S.~Fadin}{,}
\newblock Sov.~Phys.~JETP~45 (1977) 199;\\
\newblock Ya.Ya.~Balitzki and L.N.~Lipatov, Sov.~J.~Nucl.~Phys.~28
(1978) 822.

\bibitem{zeus:efl} ZEUS Collab., M. Derrick et al., Z.~Phys.~C59~(1993)~231.

\bibitem{logxpee} 
OPAL Collab., M.Akrway et al., Phys. Lett. B247 (1990) 617; \\
TASSO Collab., W.~Braunschweig et al., Z. Phys. C47 (1990) 187; \\
TASSO Collab., W.~Braunschweig et al., Z. Phys. C22 (1984) 307; \\
TOPAZ Collab., R.~Itoh et al., Phys. Lett. B345 (1995) 335.

\bibitem{seroch} S.~Lupia and W.~Ochs, Eur. Phys. J. C2 (1998) 307.

\bibitem{chlia} E.R. Boudinov, P.V. Chliapnikov and V.A. Uvarov,
Phys. Lett. B309 (1993) 210; \\
 P.V. Chliapnikov and V.A. Uvarov, Phys.Lett.B431 (1998) 430.
\bibitem{eedata}
TASSO Collab., W.~Braunschweig  et al., Z.\ Phys.\  C47
(1990) 187 ;\\
MARK II Collab., A.~Petersen  et al., Phys.\ Rev.\  D37
(1988) 1;\\
AMY Collab., Y.~K.~Li  et al., Phys.\ Rev.\  D41 (1990) 2675;\\
DELPHI Collab., P.~Abreu  et al., Phys.\ Lett.\ B311,
(1993) 408.

\bibitem{spear} MARK II Collab., J.~F.~Patrick  et al., Phys.~Rev.~Lett.
49 (1982) 1232.


\bibitem{diffeflow} ZEUS Collab., M.~Derrick et al., 
Phys. Lett. B338 (1994) 483.

\bibitem{durws} Yu.~Dokshitzer and B.~R.~Webber, discussion at 
Third UK Phenomenology Workshop on HERA Physics, Durham, UK,
 20-25 Sept 1998.\\
P.~Dixon et al., `Fragmentation Function Scaling Violations in the Breit
Frame', to appear in J.~Phys. G.


\bibitem{binnewies} J.~Binnewies et al., Z.~Phys. C65 (1995) 471.

\bibitem{photo} ZEUS Collab., M.~Derrick  et al., Z.\ Phys.\ 
 C67~(1995)~227.

\bibitem{H1photo} H1 Collab., C.~Adloff et al., DESY 98-148,
submitted to Eur. Phys. J.

\bibitem{nchee} TASSO Collab., W.~Braunschweig et al., Z. Phys. C45 (1989) 193;\\
 PLUTO Collab., Ch.~Berger et al., Phys. Lett. B95 (1980) 313; \\
OPAL Collab., P.D.~Acton et al., Z. Phys. C53 (1992) 539; \\
HRS Collab., M. Derrick et al., Phys. Rev. D34 (1986) 3304;\\
MARK1 Collab., J.L. Siegrist et al., Phys. Rev. D26 (1986) 969.

\bibitem{musgrave} M. Derrick et al., Phys. Lett. B91 (1986) 470.

\bibitem{NOMAD} NOMAD Collab., J.~Altegoer et al., Phys. Lett. B445
(1999) 439.

\bibitem{eden} P.\ Ed\'en, LU TP 98-22, hep-ph/9811229.

\bibitem{boe} Bank of England Quarterly Bulletin , February 1998, and
references therein. The two piece normal distribution is
\begin{displaymath}
\exp\left(\frac{\left(\ln(1/x_p)-\mu\right)^2}{2\Sigma^2}\left(1\pm g\right)\right),
\end{displaymath}

where $\mu\ {\rm and\ } \Sigma$ are the mode and r.m.s. of the distribution
respectively; $g$ is in the range $-1. < g < 1.$ and controls the
skewness of the distribution. If $\ln(1/x_p) \ge \mu$ then the sign in
front of $g$ is positive, otherwise it is negative.

\end{thebibliography}
\end{document}